\documentclass[]{article}
\usepackage[margin=3cm]{geometry}
\usepackage{color}
\usepackage{url}
\usepackage{geometry}
\usepackage{lipsum}
\usepackage[export]{adjustbox}
\usepackage[font=footnotesize,labelfont=bf]{caption}
\usepackage{mathtools} 
\usepackage[T1]{fontenc}
\usepackage{textcomp}
\usepackage{gensymb}
\usepackage{amsmath}
\usepackage{amssymb}
\usepackage{fancyhdr}
\usepackage[utf8]{inputenc}
\usepackage{authblk}
\usepackage{rotating}
\usepackage{comment}
\usepackage{subfig}
\usepackage{graphicx}

\title{Exact solution of stochastic gene expression models \\ with bursting, cell cycle and replication dynamics}

\author[1]{Casper H.~L.~Beentjes}
\author[2]{Ruben Perez-Carrasco}
\author[3]{Ramon Grima}
\affil[1]{Mathematical Institute, University of Oxford, UK}
\affil[2]{Department of Mathematics, University College London, UK}
\affil[3]{School of Biological Sciences, University of Edinburgh, UK}
\date{\today}

\begin{document}

\maketitle
\begin{abstract}
The bulk of stochastic gene expression models in the literature do not have an explicit description of the age of a cell within a generation and hence they cannot capture events such as cell division and DNA replication. Instead, many models incorporate cell cycle implicitly by assuming that dilution due to cell division can be described by an effective decay reaction with first-order kinetics. If it is further assumed that protein production occurs in bursts then the stationary protein distribution is a negative binomial. Here we seek to understand how accurate these implicit models are when compared with more detailed models of stochastic gene expression. We derive the exact stationary solution of the chemical master equation describing bursty protein dynamics, binomial partitioning at mitosis, age-dependent transcription dynamics including replication, and random interdivision times sampled from Erlang or more general distributions; the solution is different for single lineage and population snapshot settings. We show that protein distributions are well approximated by the solution of implicit models (a negative binomial) when the mean number of mRNAs produced per cycle is low and the cell cycle length variability is large. When these conditions are not met, the distributions are either almost bimodal or else display very flat regions near the mode and cannot be described by implicit models. We also show that for genes with low transcription rates, the size of protein noise has a strong dependence on the replication time, it is almost independent of cell cycle variability for lineage measurements and increases with cell cycle variability for population snapshot measurements. In contrast for large transcription rates, the size of protein noise is independent of replication time and increases with cell cycle variability for both lineage and population measurements. 
\end{abstract}

\section{Introduction}
It is well known that gene expression is stochastic \cite{elowitz2002stochastic}. The randomness in the time at which each reaction occurs leads to fluctuations in the molecule number of gene products such as mRNA and proteins. Hence over the past two decades there has been considerable effort devoted to constructing and solving stochastic models of gene expression \cite{kaern2005stochasticity,schnoerr2017approximation}. The exact solution of the chemical master equation (CME) describing the standard models of stochastic gene expression is currently unknown except in certain limiting cases such as when mRNA degrades much faster than protein \cite{shahrezaei2008analytical}. 

The majority of gene expression models in the literature do not have a description of cellular age and hence do not explicitly describe the cell cycle \cite{shahrezaei2008analytical,dattani2017stochastic,popovic2016geometric,bratsun2005delay,kumar2015transcriptional,tian2006stochastic,grima2012steady,thomas2014phenotypic,cao2018linear}. Rather it is assumed, following \cite{golding2005real,friedman2006linking}, that protein dilution effects due to cell division can be implicitly included via an effective first-order decay reaction. The rate of this reaction is chosen such that the half-life of protein numbers corresponds to the mean cell cycle length. This approximation is thought to be reasonable since active protein degradation timescales are considerably longer than the cell cycle time \cite{maurizi1992proteases,belle2006quantification} and hence dilution occurring during cell division is the dominant means of protein removal. 

Since these effective models do not have a description of the cell age within a cell cycle, they also cannot take into account events which happen at specific points during the cycle, e.g.\ the replication of the genome which leads to an increase of the transcription rate. The main advantage of these models is the relative ease with which they can be analytically solved, approximated and simulated. In particular, the chemical master equation of the most commonly used model of this type, which describes proteins produced in bursts whose size is sampled from the geometric distribution and protein decay via an effective first-order reaction (modelling dilution as described above), can be solved exactly leading to a negative binomial distribution (or a gamma distribution, its continuous analog) of protein numbers \cite{shahrezaei2008analytical,friedman2006linking}. 

In contrast to this implicit model, more sophisticated models have been developed during the past few years that include an explicit description of the cell cycle. Curiously, results for the case of periodic cell division were first obtained by Berg \cite{berg1978model}, about twenty years before the explosion of interest in stochastic gene expression \cite{elowitz2002stochastic}. More recently, Johnston and Jones obtained the distribution of protein numbers assuming non-bursty production, binomial partitioning at cell division and regularly spaced (periodic) cell division events \cite{johnston2015closed}. Since experimental data clearly shows that the time between two successive cell division effects is a random variable \cite{roeder2010variability,golubev2016applications}, models were also devised to study how dynamics are influenced by this extra source of randomness. Antunes and Singh \cite{antunes2015quantifying} obtained the moments of mRNA and protein numbers in a simplified model of gene expression which ignores intrinsic noise (due to the stochastic birth-death of individual molecules) and that due to binomial partitioning but takes into account noise stemming from the random timing of cell division events. Soltani et al.\ \cite{soltani2016effects,soltani2016intercellular} obtained the mean and variance of protein numbers in a considerably more detailed model for stable protein (one that is degraded only by dilution) that includes intrinsic noise, stochastic partitioning of molecules at cell division, a cell cycle that is divided in a number of phases whose duration is exponentially distributed and also can include replication. In these studies the formulae are obtained assuming single lineage measurements, i.e.\ upon cell  division, one  of  the  daughter  cells  is  followed (the other discarded) such that one obtains information about the stochastic dynamics of a cell's protein contents along an arbitrarily chosen lineage (measurements done using a mother machine such as in Ref.\ \cite{robert2018mutation}). The two major disadvantages of the latter two papers are that they do not derive results for the protein number distributions and also they do not calculate statistics in a growing population of cells, the most common experimental scenario (measurements done using flow cytometry such as \cite{newman2006single}). J\k{e}drak et al.\ \cite{jkedrak2019exactly} derived an explicit expression for the protein distribution solution of a stochastic model where protein fluctuations are treated continuously, there are no gene duplication effects, and where the cell cycle is assumed to be exponentially distributed. These results analyze the behaviour of a single lineage (as previous papers), and also for a whole proliferating population, i.e.\ both daughter cells are followed upon cell division such that one obtains information about the stochastic dynamics of a cell's protein contents across a growing population. Note that lineage and population statistics are not generally equivalent, unlike what one may assume based on the ergodic hypothesis \cite{thomas2017making}. The major limitations of the model in Ref.\ \cite{jkedrak2019exactly} are the large protein approximation implicit in the continuous approximation, the lack of DNA replication and the assumption that cell cycle duration is exponentially distributed which is contrary to experimental evidence that reveals distributions comparable to Erlang, gamma or lognormal distributions or variations thereof \cite{golubev2016applications,yates2017multi}. 

Given these two different approaches including implicit and explicit description of the cell cycle, a question arises: How well can the negative binomial distribution of implicit models describe the protein distribution of more detailed models of gene expression? This question remains unanswered because as discussed above, none of the current literature derives the protein distribution in a model that explicitly includes dilution due to stochastic partitioning of molecules at cell division, random interdivision times, and age-dependent transcription. In this paper we answer this question by deriving expressions for the distributions of proteins in models that incorporate explicit descriptions of the cell cycle. For the sake of clarity, instead of starting from the most general model, our presentation considers a set of simpler models which gradually build up to it. The three models that we study, in order of complexity, have the following properties: (i) no replication and a cell cycle of fixed duration; (ii) no replication and an Erlang distributed cell cycle duration; (iii) age-dependent transcriptional dynamics including replication and a cell cycle described by a number of phases each of which has an exponentially distributed duration (hypoexponential distribution). All the models consider proteins that are produced in bursts \cite{yu2006probing}, degraded only via dilution (stable proteins \cite{friedman2006linking}) and assume binomial partitioning of proteins at cell division \cite{berg1978model}. We study the relationships between the solutions of all models for both single lineage and population snapshot statistics, and identify conditions under which the protein distributions can be well approximated by the negative binomial solution of the conventional model of gene expression with an implicit cell cycle description.  

\section{Model I: Stochastic gene expression with an implicit description of the cell cycle}\label{sec:ModelI}

It is well known that under the assumptions that mRNA degrades much faster than protein and that promoter switching is also much faster than protein decay, the stochastic dynamics of protein $P$ can be effectively described by the reaction scheme \cite{shahrezaei2008analytical}
\begin{equation}
\label{ModelI}
G \xrightarrow{r} G + m P,\quad P \xrightarrow{d} \emptyset,
\end{equation}
where $G$ denotes a single gene copy, $r$ is the effective burst production rate and $d$ is the protein degradation rate. Note that protein is produced in bursts of size $m$, which follows a geometric distribution (in accordance with experiments \cite{yu2006probing}), i.e.\ $m\sim \text{Geom}(p)$ with a mean burst size $\alpha = (1-p)/p=h/d_m$ where $h$ is the protein translation rate and $d_m$ is the mRNA degradation rate. This burstiness implicitly describes the mRNA dynamics since a burst in protein expression occurs due to rapid translation of proteins by a single short-lived mRNA. Hence within this context, the parameter $r$ is also the same as the effective mRNA transcription rate which is given by $r = \rho_u \sigma_u / (\sigma_b + \sigma_u)$ where $\rho_u$ is the mRNA transcription rate, $\sigma_b$ is the rate of switching from the active state to the inactive state and $\sigma_u$ is the rate of switching from the inactive state to the active state. 

Note that this model has been rigorously derived from a three-stage model of gene expression that does not take the cell cycle explicitly into account \cite{shahrezaei2008analytical}. However, it is commonly assumed that the protein degradation reaction effectively models the dilution which occurs due to binomial partitioning at cell division. The question then is how we should choose the effective protein degradation rate. A simple argument is as follows. Since the protein decays exponentially via an effective first-order reaction, its half-life is $t_d = \log(2)/d$; we also know that the protein concentration is on average halved at cell division due to binomial partitioning and hence $t_d = T$, where $T$ is the cell cycle length. Hence it follows that the effective protein degradation rate should be chosen as $d=\log(2)/T$. The effective model given by reaction scheme \eqref{ModelI} and $d$ chosen as aforementioned is one of the standard models of gene expression in the literature \cite{friedman2006linking}.      

In Appendix A.1, we show that the effective model described above provides an accurate description of the mean number of proteins in a three-stage model of gene expression with explicit mRNA and protein dynamics, binomial partitioning and fixed cell cycle length $T$, provided (i) the mRNA degrades much faster than protein; (ii) promoter switching is much faster than protein decay; (iii) mRNA degrades on a much shorter timescale than the cell cycle length $T$; (iv) the mean number of proteins is calculated from population measurements. Note that if instead the mean number of proteins is calculated from single lineage measurements then we arrive at model \eqref{ModelI} but with effective protein degradation rate given by $d=(2/3)/T$ (see Appendix A). Note that effective degradation rates derived for lineage data are slightly smaller than their population snapshot equivalent, because $2/3<\log 2$. This discrepancy stems from the fact that the mean number of proteins calculated from population measurements is smaller than the mean number of proteins calculated in single lineage measurements. This is because in population snapshots, all cells are tracked and hence due to a doubling of the number of cells at cell division, there is a bias towards observing young cells with small protein counts. In contrast, in lineage measurements since only one cell is tracked, the probability of observing a cell of any age is the same and hence the protein counts on average are higher than in population measurements. A detailed discussion of the difference between these two types of measurement can be found in \cite{thomas2017making}.

The CME of Model I whose reactions are given by \eqref{ModelI} is straightforward to solve using the method of generating functions. In steady-state, its solution reads
\begin{equation}
\label{genNB}
G(z) = \biggl(\frac{1}{1-\alpha(z-1)}\biggr)^{\beta},
\end{equation}
where $G(z)=\sum_n z^n P(n)$ is the probability generating function (PGF), $P(n)$ is the steady-state protein distribution of protein number $n$ and $\beta = r/d$. Note that $\beta$ is the average number of mRNA molecules produced in the protein life-time. In the case of stable proteins (which degrade only by dilution) it follows from our previous results for effective degradation rates that  $\beta=3y/2$ for lineage measurements and $\beta=y/\log2$ for population snapshot measurements, where we have defined $y=rT$ as the average number of mRNA produced in a cell cycle. The distribution can be obtained using $P(n)=(1/n!)dG^n/dz^n|_{z=0}$ which leads to a negative binomial $\text{NB}(\beta,\alpha/(1+\alpha))$. 

\section{Model II: Stochastic gene expression with explicit modelling of a fixed length cell cycle}

Next we consider a model where protein production occurs in bursts as in Model I but there is no effective first-order reaction modelling protein degradation. Instead we explicitly model binomial partitioning of the proteins at cell division. The major assumption of this model is that cell division occurs at regular time intervals of length $T$. This is often referred to as a ``timer'' mechanism and has been found in certain types of cells, e.g.\ early frog embryos \cite{wang2000transition}. In what follows we will find an exact steady-state solution of the CME for this model and compare it with that of Model I.  

Let $t\in[0,T]$ be the age of a given cell, namely $t = 0$ corresponds to its birth and $t=T$ corresponds to the time at which it divides into two. The CME describing bursty protein expression and no active degradation in a cell is given by
\begin{equation}
\label{CMEModelII}
\frac{dP^j(n,t)}{dt} = r\sum_{m=0}^\infty P^j(n-m,t)Q(m) - r P^j(n,t) \sum_{m=0}^\infty Q(m),  
\end{equation}
where $P^j(n,t)$ is the probability that at cell age $t$ in generation $j$ there are $n$ proteins observed. Here $Q(m)=p(1-p)^m$ with $\alpha=(1-p)/p$ is the geometric distribution with mean $\alpha$. Note that each time cell division occurs, the generation number $j$ is increased by one. The PGF equation corresponding to the CME is
\begin{equation}\label{PGFpdeModelII}
\frac{\partial G^j(z,t)}{\partial t} = -rG^j(z,t)\biggl(1 - \frac{1}{1+\alpha(1-z)} \biggr),
\end{equation}
which has a time-dependent solution
\begin{equation}
\label{genfneq}
G^j(z,t) = F^j(z) \exp\biggl(\frac{-\alpha r t (1-z)}{1 + \alpha(1-z)}\biggr).
\end{equation}
Note that $F^j(z) = \sum_n z^n P^j(n,0)$ namely the PGF corresponding to the protein distribution at cell birth in generation $j$.

Introducing binomial partitioning at mitosis leads to a simple relationship between the protein distribution at cell division of a cell in generation $j$ and the distribution observed at the birth of the daughter cell in generation $j+1$
\begin{equation}
\label{binomModII}
P^{j+1}(n,0)=\sum_{i=0}^{\infty} \binom{i}{n} 2^{-i} P^j(i,T),
\end{equation}
which implies for the PGF
\begin{equation}
\label{partition1}
F^{j+1}(z)=G^{j+1}(z,0)=G^j\biggl(\frac{1+z}{2},T\biggr).    
\end{equation}
Note that in Eq.\ \eqref{binomModII} we used the convention that $i$ choose $n$ equals zero when $n < i$. In this case we cannot impose steady-state as in Model I because cell division occurs at regular time intervals. Rather we consider cyclo-stationary conditions which are achieved when the probability that a cell of age $t$ has a given number of proteins is independent of which generation it belongs to, i.e.\ the superscript $j$ in Eqs.\ \eqref{CMEModelII}-\eqref{partition1} can be ignored. Hence substituting Eq.\ \eqref{partition1} in Eq.\ \eqref{genfneq} we obtain
\begin{equation}
\label{genfneqmain}
G(z,t) = G \biggl(\frac{1+z}{2},T\biggr) \exp\biggl(\frac{-\alpha r t (1-z)}{1 + \alpha(1-z)}\biggr).
\end{equation}

Next we proceed to solve Eq.\ \eqref{genfneqmain} by substituting $t=T$ in this equation to obtain
\begin{equation}
\label{gnfncelldiv}
G(z,T) = G \biggl(\frac{1+z}{2},T\biggr) f(z),
\end{equation}
where $f(z) = \exp(-\alpha r T (1-z)/(1 + \alpha(1-z))$. Eq.\ \eqref{gnfncelldiv} can be solved by iteration as follows
\begin{align}
\notag
G(z,T) &= f(z) f\biggl(\frac{1+z}{2}\biggr) G\biggl(\frac{3+z}{4},T\biggr) , \\ \notag &=f(z) f\biggl(\frac{1+z}{2}\biggr) f\biggl(\frac{3+z}{4}\biggr) G\biggl(\frac{7+z}{8},T\biggr) , \\ \label{eqGzT} &= G(1,T) \prod_{s=0}^\infty f\biggl(\frac{2^s+z-1}{2^s}\biggr) = \prod_{s=0}^\infty f\biggl(\frac{2^s+z-1}{2^s}\biggr).
\end{align}
Note that here we used $G(1,T)=1$ which follows from the normalisation of the distribution. Substituting Eq.\ \eqref{eqGzT} in Eq.\ \eqref{gnfncelldiv} we obtain $G((1+z)/2,T)$ which after substituting in Eq.\ \eqref{genfneqmain} leads us to an explicit solution of the generating function for Model II in cyclo-stationary conditions
\begin{align}
\label{solgenfnModelII}
\notag
G(z,t) &= \left(\prod_{s=0}^\infty f\biggl(1+\frac{z-1}{2^{s+1}}\biggr)\right) \exp\biggl(\frac{-\alpha rt (1-z)}{1 + \alpha(1-z)}\biggr), \\
&= \exp \biggl(\frac{r x t}{1 - x} + rxT \sum_{s=0}^\infty \frac{1}{2^{1+s} - x}\biggr),
\end{align}
where in the last line we used the definition of the function $f$ and the definition $x = \alpha(z - 1)$. Note that the sum over $s$ in the argument of the exponent can be written in terms of the $q$-digamma function.

The solution we have computed corresponds to the PGF of the protein distribution computed from an ensemble of identical cells all of which are at the same cell age $t$. However, distributions are often calculated from experimental measurements of time traces of the fluorescent protein molecules along a cell lineage or else from population snapshots. Considering the single lineage case, the corresponding PGF is given by a time-average of the generating function calculated earlier
\begin{align}
\label{genfnsingle}
G_{s}(z)&=\int_{0}^T \frac{1}{T} G(z,t) dt \notag \\ 
&=\frac{x-1}{x y}\biggl(\exp\biggl(\frac{x y}{x-1}\biggr)-1\biggr)\exp\biggl(x y\sum_{s=0}^\infty \frac{1}{2^s-x}\biggr)
\end{align}
where we have used $y=rT$ as we did for Model I. Note that the subscript $s$ will henceforth be used to denote single lineage measurement. Note also that the time-average is computed since the probability of observing cells of any age is uniform in lineage measurements. Comparing this PGF to that of Model I, i.e.\ Eq.\ \eqref{genNB}, it is clear that in Model II the protein distribution is generally not equal to the negative binomial distribution of Model I.  

To understand the differences between these two distributions, we next use the PGF's given by Eq.\ \eqref{genNB} (with $\beta = 3y/2$) and Eq.\ \eqref{genfnsingle} to compute the mean $\langle n \rangle$ and variance $\sigma^2$ in stationary conditions
\begin{align}
\label{eqmeanNB1a}
\langle n \rangle_{NB,s}&=\frac{3}{2}\alpha y ,\\ 
\sigma^2_{NB,s} &= \frac{3}{2}\alpha y + \frac{3}{2}\alpha^2 y= 1.5\alpha y + 1.5\alpha^2y, \\
\langle n \rangle_{s}&=\frac{3}{2}\alpha y, \\ \label{eqvarM2a}
 \sigma^2_{s} &=\frac{3}{2}\alpha y + \alpha^2\left(\frac{5y}{3} + \frac{y^2}{12}\right) \approx 1.5\alpha y + 1.67\alpha^2y + 0.08\alpha^2y^2.
\end{align}
It is clear that while the mean of the two distributions is the same, the variances are generally different. In Appendix A, we clarify the origin of this discrepancy. In particular, Model I, under certain conditions described earlier, can match the mean number of proteins in a three-stage model of gene expression with explicit mRNA and protein dynamics, binomial partitioning and fixed cell cycle length $T$ whereas Model II can match the full PGF of the three-stage model under the same conditions (see in particular Appendix A.2). The variance of Model II is always greater than that of Model I ($\sigma^2_{s} > \sigma^2_{NB,s}$). Furthermore, whilst the two variances are both quadratic in $\alpha$, $\sigma^2_{NB,s}$ is linear in $y$ while $\sigma^2_{s}$ is quadratic in $y$. The relative error between the two variances computed as $(\sigma^2_{s}-\sigma^2_{NB,s})/\sigma^2_{s}$ increases monotonically with $\alpha$ and $y$ but is mostly determined by the value of $y$, i.e.\ the average number of mRNA produced in a cell cycle, as illustrated in Fig.\ 1a. Expressions for the skewness squared can also be easily derived from the PGF
\begin{align}
\label{sk1}
S^2_{NB,s} &=\frac{2(2 \alpha+1)^2 }{3 \alpha (\alpha+1) y}, \\
\label{sk2}
S^2_{s}&=\frac{108 \left(2 \alpha^2 (7 y+54)+7 \alpha (y+20)+42\right)^2}{49 \alpha y (\alpha
   (y+20)+18)^3}. 
\end{align}
For small $y$, both of these expressions are proportional to $1/y$ while for large $y$, we have $S^2_{NB,s} \propto 1/y$ while $S^2_{s} \propto 1/y^2$, i.e.\ for large enough $y$ Model II will predict a less skewed distribution than Model I. Generally the skewness of the distribution of Model II can be larger or smaller than that of Model I depending on the values of $\alpha$ and $y$ (Fig.\ 1b). 

To get a fuller picture of the differences between the two models (assuming lineage measurements) we plot in Fig.\ 2 the distributions for various values of $y$ whilst keeping the value of $\alpha$ fixed. Note that the distribution of protein numbers for Model II is constructed by first expanding the generating function Eq. \eqref{genfnsingle} as a Taylor series using a symbolic computation software and then $P(n)$ is simply given by the $n^{\text{th}}$ coefficient of this series.
The theoretical predictions for Model II are also verified by means of the stochastic simulation algorithm (SSA, see Appendix B for a full description of the algorithm). The negative binomial distribution of Model I (red line) is a good approximation of the distribution of Model II (black line) for small $y$ but clearly is inappropriate for large $y$; it can also be shown that the difference between distributions becomes more pronounced if $\alpha$ is increased. 

A visual inspection of the distribution of Model II (black line) leads one to believe that one can likely fit well an effective negative binomial for the cases shown in Fig.\ 2(a)-(c) but not for (d). This intuition is verified in Fig.\ 2 by plotting an effective negative binomial of the same mean and variance (green open circles) as the distribution of Model II; in Fig.\ 2d, the distribution is considerably flatter near the mode than the fitted negative binomial. Note that the effective negative binomial is given by $\text{NB}(z_1,z_2)$ where
\begin{align}
\label{z1est}
z_1 &=\frac{27 y}{y+20},\\
\label{z2est}
z_2 &=\frac{\alpha (y+20)}{\alpha (y+20)+18}.
\end{align}

Since Model I has a negative binomial solution, it follows that through renormalisation of its parameters,  it can be matched to the effective negative binomial for Model II. If the renormalised parameters for Model I are $y_e$ and $\alpha_e$, then its solution is $\text{NB}(z_1,z_2)$ where $z_1=3 y_e/2,z_2=\alpha_e/(1+\alpha_e)$. Equating $z_1, z_2$ to those in Eqs.\ \eqref{z1est}-\eqref{z2est}, we obtain  
\begin{align}
\alpha_e &=  \alpha \left(\frac{y+20}{18}\right), \\
y_e &= y \left(\frac{18}{y+20}\right).
\end{align}
Note that renormalised Model I is the same as the green open circles shown in Fig.\ 2 which is a much better approximation to Model II (black line) than the original Model I (red line). From these equations, it is also clear that if parameters had to be estimated from experimental data (with low cell cycle duration variability) using Model I, then the estimated mean burst size $\alpha_e$ overestimates the true value $\alpha$, while the estimated mean number of mRNA per cell cycle $y_e$ underestimates the true value $y$.

In this section we have so far focused on the distributions for single lineage measurements. The distribution of protein numbers for population snapshots can also be derived and instead of Eq.\ \eqref{genfnsingle} we then have
\begin{align}
\label{genfnpop}
G_{p}(z)&=\int_{0}^T \frac{2^{1-t/T}\log(2)}{T} G(z,t) dt, \notag \\ &= \frac{(x-1) \log 2 }{x y+(x-1) \log 2}\left( \exp\left(\frac{x y+(x-1) \log 2}{x-1}\right)-1\right)\exp\left( xy \sum_{s=0}^{\infty}\frac{1}{2^s-x}\right),
\end{align}
where the subscript $p$ will be used to denote population snapshot measurements from hereon. Note that we used the fact that when interdivision times are regularly spaced in time, the probability of observing a cell of age $t \in[0,T]$ is $2^{1-t/T}\log2/T$ for population measurements \cite{berg1978model} (see also Appendix D for a derivation of the latter). The PGF of Model I has already been calculated for the population scenario in Section 2 and was found to be given by Eq.\ \eqref{genNB} (with $\beta = y/\log(2)$). For the parameters used in Fig.\ 2, the protein distributions for population data for Models I, II are found to be close to those calculated for lineage data and hence we do not show them. The mean and variance of the protein numbers of Models I and II are now given by
\begin{align}
 \langle n \rangle_{NB,p} &= \frac{\alpha y}{\log(2)}, \\
 \sigma^2_{NB,p} &=\frac{\alpha (1+\alpha) y}{\log(2)} \approx 1.44 \alpha y + 1.44 \alpha^2 y, \\
 \langle n \rangle_p &= \frac{\alpha y}{\log(2)}, \\
    \sigma^2_p &= \frac{1}{\log 2}\alpha y + \frac{6 - 4\log 2}{3\log 2}\alpha^2 y + \frac{1-\log^2 2}{\log^2 2} \alpha^2 y^2 \approx 1.44\alpha y +1.55\alpha^2 y + 0.08\alpha^2 y^2.
\end{align}
These equations are the population equivalent of Eqs.\ \eqref{eqmeanNB1a}-\eqref{eqvarM2a} and the same observations we made earlier regarding the comparison of the moments of Model I and Model II for single lineage data are also seen to hold for population data. Note also that generally we can state $\langle n \rangle_{p} < \langle n \rangle_{s}$ and $\sigma^2_{p} < \sigma^2_{s}$, which can be explained by the enhanced probability of observing younger cells (and hence having a smaller protein content) in population measurements (as mentioned in Section 2). 

Summarizing, our results in this section imply that the effective degradation reaction in Model I cannot effectively account for dilution via binomial partitioning. Generally the models agree on the mean number of proteins in stationary conditions but not on the higher-order moments. The discrepancies are particularly obvious whenever $y$ is greater than a few tens. The variance of Model I is always less than that of Model II but the skewness of Model I can be greater or smaller than that of Model II. We have also shown that the protein distribution of Model II can be well approximated by an effective negative binomial distribution only if $y$, the mean number of mRNAs produced in a cell cycle, is small. In this case, it is possible to renormalise the parameters in Model I so that its solution approximates that of Model II well.
    
\section{Model III: Stochastic gene expression with explicit modelling of an Erlang distributed cell cycle length}

Next we consider a more complex and realistic model of bursty gene expression, namely one that includes cell cycle length variability. Such variability could for example originate in cell types where the cell growth rate is stochastic and cell division is triggered when the volume of a cell exceeds its volume at birth by a certain fixed amount (also called an ``adder'' mechanism, see for example \cite{cadart2018size}). As we shall see, this requires a very different master equation description than the previous models. Specifically, the model has the following properties: (i) the cell cycle is divided in $N$ phases where the duration of each phase is exponentially distributed with parameter $k$. It then follows that the cell cycle length distribution is Erlang, the mean cell cycle time is $T = N/k$ and the coefficient of variation of the cell cycle duration is $1/\sqrt{N}$. (ii) proteins are produced at a rate $r$ and in geometrically distributed burst sizes with mean $\alpha$. (iii) cell division occurs instantaneously after the end of the $N^{\text{th}}$ phase which leads to binomial partitioning of proteins between mother and daughter cells. Note that an Erlang distribution provides a good fit to some of the measured cell cycle length distributions \cite{antunes2015quantifying,yates2017multi}.

Let $P_i(n,t)$ be the probability that the cell cycle is in phase $i$ at time $t$ and that there are $n$ protein molecules. We shall here ignore the generation number since in steady-state conditions this does not matter. It then follows that the master equation describing the above model is given by
\begin{align}
\label{eq1E}
\frac{d P_1(n,t)}{dt} = -&k P_1(n,t) + k P_N'(n,t) + \nonumber \\ &r\sum_{m=0}^\infty P_1(n-m,t)Q(m) - r P_1(n,t) \sum_{m=0}^\infty Q(m),\\
\label{eq2E}
\frac{d P_i(n,t)}{dt} = -&k P_i(n,t) + k P_{i-1}(n,t) + \nonumber \\ &r\sum_{m=0}^\infty P_i(n-m,t)Q(m) - r P_i(n,t) \sum_{m=0}^\infty Q(m), \quad i \in [2,N]
\end{align}    
where $Q(m)=p(1-p)^m$ is the geometric distribution with mean $\alpha=(1-p)/p$. The first term in these equations models exit from the present cell cycle phase into the next phase, the second term models the entry into the present cell cycle phase from the previous one, and the third term models bursty protein production. Note that binomial partitioning during cell division is explicitly taken into account by the second term of Eq.\ (\ref{eq1E}). In particular this process implies
\begin{align}
\label{Pbinopart}
P_N'(n,t) = \sum_{m=0}^{\infty} \binom{m}{n} 2^{-m} P_N(m,t),
\end{align} 
where we take the convention $m$ choose $n$ equals zero when $n < m$. The PGF equations corresponding to the CME Eq.\ \eqref{eq1E}-\eqref{eq2E} are given by
\begin{align}
\label{eqGE1}
\frac{\partial G_{1,s}(z)}{\partial t} &= -kG_{1,s}(z) + kG_{N,s}\biggl(\frac{1+z}{2} \bigg) -rG_{1,s}(z)\biggl(1 - \frac{1}{1+\alpha(1-z)} \biggr), \\ 
\label{eqGE2} 
\frac{\partial G_{i,s}(z)}{\partial t} &= -kG_{i,s}(z) + kG_{i-1,s}(z) - rG_{i,s}(z)\biggl(1 - \frac{1}{1+\alpha(1-z)} \biggr), \ i \in [2,N],
\end{align}
where we have suppressed the time dependence for convenience. Note that while for Model II, the cyclo-stationary condition meant that the protein distribution at a given cell age is independent of generation number, for Model III the condition means that the protein number at a given cell cycle phase is independent of the generation number. Hence we can set Eq.\ \eqref{eqGE2} to zero and solve recursively for $G_{i,s}(z)$ to obtain
\begin{align}
\label{eqGrecE}
G_{i,s}(z) = \biggl(\frac{k(x-1)}{k(x-1)+rx}\biggr)^{i-1} G_{1,s}(z), \quad i \in [2,N].
\end{align}
Substituting Eq.\ \eqref{eqGrecE} with $i = N$ in Eq.\ \eqref{eqGE1} with the left hand side equal to zero, we obtain
\begin{align}
G_{N,s}\biggl(\frac{1+z}{2} \bigg) &= \frac{k(x-1)+rx}{k(x-1)} G_{1,s}(z), \notag \\&=\biggl(\frac{k(x-1)}{k(x-1)+rx}\biggr)^{-N} G_{N,s}(z).
\end{align}
Following the same method of solution as used for solving Eq.\ \eqref{gnfncelldiv}, we obtain
\begin{align}
\label{genfnNfinal_aa}
G_{N,s}(z) = \prod_{s=0}^\infty \left(1-\frac{r x}{x (k+r)-2^s k}\right)^N,
\end{align}
where we used the normalisation condition for the conditional distribution in each phase, i.e.\ $G_i(1)=1$. Using Eq.\ \eqref{eqGrecE} and Eq.\ \eqref{genfnNfinal_aa} we obtain the PGF for the conditional protein distribution in cell phase $i$
\begin{align}
G_{i,s}(z) = \left(1 + \frac{rx}{k(x-1)}\right)^{N-i} \prod_{s=0}^\infty \left(1-\frac{r x}{x (k+r)-2^s k}\right)^N, \quad i \in [1,N].
\end{align}
Since we are considering the case of single lineage measurements, we must average the PGF over all cell phases by marginalising out the phase in which a cell is at observation time. Note that in order to do this we need an expression for $\Pi_{i,s}$, the probability that the cell is in phase $i$ at observation in a single lineage measurement. For our case of $N$ identically exponentially distributed phases it can be easily shown that $\Pi_{i,s}=N^{-1}$, reflecting that every phase is equally likely to be observed. Using this we derive the PGF for the protein distribution for lineage measurements
\begin{align}
\label{genfnErlangfinal1}
G_s^E(z) = \sum_{i=1}^N \Pi_{i,s} G_{i,s}(z) = \frac{x-1}{xy} \left(\left(1 + \frac{xy}{N(x-1)}\right)^N - 1\right) \prod_{s=0}^\infty \left(1-\frac{xy}{x (N+y)x-2^s N}\right)^N,   
\end{align}
where we used that $y = rT$ and the mean cell cycle length $T = N/k$. Note that the superscript {\it{E}} denotes an Erlang distributed cell cycle length. This solution can be conveniently written in terms of exponential functions yielding
\begin{align}
\label{genfnErlangfinal2}
G_s^E(z) = \frac{x-1}{xy} \left(\exp \left(N \log \left(1 + \frac{xy}{N(x-1)}\right) \right) - 1 \right) \exp \left(N \sum_{s=0}^\infty \log \left(1 - \frac{xy}{(N+y)x - 2^s N} \right)\right).
\end{align}
Note that the argument of the log is always positive because $x = \alpha(z-1) \le 0$ due to $z \le 1$. Note also that in the limit of large $N$ (at constant $T$), the Erlang distribution describing the cell cycle length tends to a delta function centered on $T$, i.e.\ a cell cycle of fixed length. It is straightforward to show by a series expansion in $1/N$ that in the limit of large $N$, Eq.\ \eqref{genfnErlangfinal2} converges to Eq.\ \eqref{genfnsingle}, i.e.\ in the limit of small cell cycle length variability, Model III converges to Model II. The mean and variance of the protein number distribution in steady-state conditions can be straightforwardly computed from the PGF
\begin{align}
\label{meanErlang}
\langle n \rangle_{s}^E&=\frac{\alpha y (3N+1)}{2 N}, \\
\label{varErlang}
(\sigma^E_{s})^2 &= \frac{\alpha y (3 N+1)}{2 N} + \alpha^2 \left(\frac{(N (N+10)+5) y^2}{12 N^2}+\frac{(5 N+3) y}{3
   N}\right).
\end{align}
 
As expected, the variance of Model III is always larger than that of Model II; the mean of Model III is slightly larger than that of Model II but the difference can be ignored in most cases of interest. Both the mean and variance are monotonic decreasing functions of $N$ and hence they are bounded from above by the moments evaluated for $N = 1$, i.e.\ an exponentially distributed cell cycle length.

The differences in the protein number distributions predicted by Models II and III for lineage observations are illustrated in Fig.\ 3a-c. There we show the excellent agreement between the theoretical expressions and stochastic simulations for both small $y$ (Fig.\ 3a) and large $y$ (Fig.\ 3b,c). While the differences between Model I and Model II are due to binomial partitioning, the differences between Model II and Model III are due to cell cycle length variability. We find that typically for $N > 20$ the differences between Models II and III are small; the differences are at their largest when the cell cycle length is exponentially distributed, i.e.\ $N = 1$.  
Previously we saw how for Model II the negative binomial was not a good fit for the distribution when $y$ was large. However as can be appreciated from Fig.\ 3b and 3c, the fit becomes better when we consider cell cycle length variability: the deviations from negative binomial, which manifest in the flattish region around the mode, become less visible as $N$ decreases. Note also that the deviations from negative binomial are largely unaffected by the value of the mean burst size (increasing $\alpha$ by ten times in Fig. 3 causes the protein distributions to move right and their height to be rescaled but their shape remains practically unaltered).  

Since the solution of Model I is generally a negative binomial, it follows that we can renormalise the parameters of this model such that its protein distribution provides a good match to the distribution of Model III when the cell cycle length variability is sufficiently high. Equating the mean and variance of a negative binomial $\text{NB}(z_1,z_2)$ to Eqs.\ \eqref{meanErlang} and \eqref{varErlang} we find
\begin{align}
z_1 &= \frac{3 (3 N+1)^2 y}{(N (N+10)+5) y+4 N (5 N+3)}, \\ z_2 &= \left(\frac{6 N (3 N+1)}{\alpha (N (N+10)+5) y+4 \alpha N (5 N+3)}+1\right)^{-1}.    
\end{align}
Equating these two parameters $(z_1,z_2)$ to those of Model I with renormalised parameters ($3 y_e/2$,$\alpha_e/(1+\alpha_e)$), we obtain the relationship between the actual and renormalised parameters
\begin{align}
\label{renormparamA}
\alpha_e &= \frac{z_2}{1-z_2} = \alpha\left(1 + \frac{N+3}{3(3N+1)} + y\left(\frac{N^2+10N+5}{6N(3N+1)} \right)\right), \\ \label{renormparamB}
y_e &= \frac{2z_1}{3} = y\left( \frac{2(3N+1)^2}{(N^2+10N + 5)y + 4N(5N+3)}\right).
\end{align}

It also follows from these formulae that if we had to fit a negative binomial to experimental data from cells with an Erlang distributed cell cycle length (data consistent with Model III) and estimate the parameters using Model I, then this will lead to an over-estimate for the mean burst size and an under-estimate for the mean number of mRNAs per cycle (and hence for the transcription rate). The errors increase with decreasing $N$ and hence with increasing cell cycle duration variability. 

We have here focused on the distributions for single lineage measurements. The distributions of protein numbers for population snapshots can also be derived. Due to the rather more complex analysis involved, the derivation is presented in Appendix \ref{sec:popgenfn}. Here we simply state the equivalent of Eq.\ \eqref{genfnErlangfinal1} for population measurements
\begin{align}
\label{genfnErlangfinal1pop}
G_p^E(z) = \left(2^{\frac{1}{N}}-1\right) N (x-1)  \frac{\left( \frac{x y}{N
   (x-1)}+2^{\frac{1}{N}}\right)^N - 1}{\left(2^{\frac{1}{N}}-1\right) N (x-1)+x y} \ \underset{s=0}{\overset{\infty }{\prod }}\left(1-\frac{x y}{x (N 2^{1/N}+y)-N 2^{1/N} 2^s}\right)^{N}.
\end{align}
The protein distributions corresponding to this PGF are shown in Fig.\ 3(d)-(f) where they are also compared with those of Model II. Note that given the same parameters, the protein distributions for lineage and populations observations are considerably different. These differences become more appreciable with increasing $y$ and decreasing $N$. While the increase in cell cycle length variability (through decreasing $N$) results in little changes to the mode of the lineage distribution, it causes the mode of the population distribution to shift to the left. However there are also qualitative similarities, namely that in both cases the deviations from negative binomial are maximal for small cell cycle length variability (large $N$) and large $y$. Similar to what we previously did for lineage observations (see Eqs.\ \eqref{renormparamA} and \eqref{renormparamB}), from the equations for the mean and variance for population snapshots (see Appendix \ref{sec:popgenfn}), it is also possible to calculate the renormalised parameters in Model I such that it provides a good negative binomial approximation to the population distribution of Model III when there is sufficient cell cycle length variability. 

Summarizing in this section we have studied a model (Model III) of bursty gene expression with an Erlang distributed cell cycle length. This model recovers Model II in the limit of small cell cycle length variability. Also the presence of sufficient cell cycle length variability is found to lift the deviations from negative binomial observed for Model II; in this case by an appropriate renormalisation of the parameters, Model I can describe the distribution predicted by Model III well. The mean and variance of protein numbers calculated from a similar model (assuming lineage observations) have been reported previously \cite{soltani2016effects}.

\section{Model IV: Stochastic gene expression with hypoexponential cell cycle length distribution and age-dependent transcription}

We next consider a more general version of Model III: (i) the time spent in phase $i$ of the cell cycle is exponentially distributed with parameter $k_i$, which implies that the cell cycle length distribution is hypoexponential. (ii) the transcription rate and burst size are age dependent, i.e.\ they are $r_i$ and $\alpha_i$ in phase $i$ respectively. Note that the Erlang distribution is a special case of the hypoexponential distribution and hence the use of this distribution, in principle, allows more flexibility in fitting experimental cell cycle distributions. Note also that modelling the transcription rate as age dependent enables us to capture replication, hence considerably extending the realism of our model. The master equation describing the above model is a generalisation of that for Model III and is given by
\begin{align}
\label{eq1H}
\frac{d P_1(n,t)}{dt} = -&k_1 P_1(n,t) + k_N P_N'(n,t) + \nonumber \\ &r_1\sum_{m=0}^\infty P_1(n-m,t)Q_1(m) - r_1 P_1(n,t) \sum_{m=0}^\infty Q_1(m),\\
\label{eq2H}
\frac{d P_i(n,t)}{dt} = -&k_i P_i(n,t) + k_{i-1} P_{i-1}(n,t) + \nonumber \\ &\sum_{m=0}^\infty r_i P_i(n-m,t)Q_i(m) - r_i P_i(n,t) \sum_{m=0}^\infty Q_i(m), \ i \in [2,N],
\end{align}    
where $Q_i(m)=p_i(1-p_i)^m$ is the geometric distribution with mean $\alpha_i=(1-p_i)/p_i$. Note that $P_N'(n,t)$ is defined as before using Eq.\ \eqref{Pbinopart}. The corresponding PGF equations are given by
\begin{align}
\label{eqGE1H}
\frac{\partial G_{1,s}(z)}{\partial t} &= -k_1G_{1,s}(z) + k_N G_{N,s}\biggl(\frac{1+z}{2} \bigg) -r_1G_{1,s}(z)\biggl(1 - \frac{1}{1+\alpha_1(1-z)} \biggr), \\ \label{eqGE2H} \frac{\partial G_{i,s}(z)}{\partial t} &= -k_i G_{i,s}(z) + k_{i-1} G_{i-1,s}(z) - r_i G_{i,s}(z)\biggl(1 - \frac{1}{1+\alpha_i(1-z)} \biggr), \ i \in [2,N],
\end{align}
where we have suppressed the time dependence for convenience. Setting Eq.\ \eqref{eqGE2H} to zero (steady-state conditions) and solving recursively for $G_{i,s}(z)$ we obtain
\begin{align}
\label{eqGrecH}
G_{i,s}(z) = w_{i,s}(z) G_{1,s}(z),
\end{align}
where 
\begin{align}
w_{i,s}(z) = 
\begin{cases}
      1 & i=1,\\
      \prod_{j=2}^i \frac{k_{j-1}(x_j-1)}{k_j(x_j-1)+r_jx_j} & i \in [2,N].
\end{cases}
\end{align}
Note that here we defined $x_j=\alpha_j(z-1)$. Substituting Eq.\ \eqref{eqGrecH} with $i = N$ in Eq.\ \eqref{eqGE1H} with the left hand side equal to zero, we obtain
\begin{align}\label{eq:genfnNrecurrence}
G_{N,s}\biggl(\frac{1+z}{2} \bigg) &=G_{N,s}(z) \left(\frac{k_1(x_1-1)+r_1 x_1}{k_N(x_1-1)w_N(z)}\right).
\end{align}
Following the same method of solution as used for solving Eq.\ \eqref{gnfncelldiv}, we obtain
\begin{align}
\label{genfnNfinal}
G_{N,s}(z) = \prod_{s=0}^\infty \prod_{j=1}^N \left(1-\frac{r_j x_j}{x_j (k_j+r_j)-2^s k_j}\right),
\end{align}
where again we used the normalisation condition for the conditional distribution in each phase, i.e.\ $G_{i,s}(1)=1$. Hence finally by means of Eq.\ \eqref{eqGrecH} and Eq.\ \eqref{genfnNfinal}, we can write an equation for the PGF of the distribution assuming single lineage measurements
\begin{align}
\label{finalgenfnhypoexp_i}
G_{i,s}(z)&=\frac{ w_{i,s}(z)}{w_{N,s}(z)}G_{N,s}(z),\\
\label{finalgenfnhypoexp}
G_s^H(z)&=\sum_{i=1}^N \Pi_{i,s} G_{i,s}(z), 
\end{align}
where again we let $\Pi_{i,s}$ denote the probability of observing the cell in phase $i$ in a lineage measurement, which can be shown to be given by $\Pi_{i,s}=k_i^{-1}/(\sum_j k_j^{-1})$. Note the superscript $H$ (standing for hypoexponential) is to distinguish this PGF from the one calculated for Model III using the Erlang distribution.  

\subsection{Modelling Replication}

We next use this theory to understand the effects of gene replication on stochastic gene expression. One of the simplest models of this process assumes (i) the transcription rate to be a constant $r$ before replication, doubling to $2r$ right after replication (the doubling in transcription rate is due to the doubling in gene copy number during replication). (ii) the cell cycle length and the replication time are Erlang distributed. This implies the special case where $k_i = k$ and $\alpha_i=\alpha$ for all $i$, and $r_i = r, \ i \in [1,M]$ and $r_i = 2r, \ i \in [M+1,N]$, where $M$ is the cell cycle phase after which replication occurs. Substituting these values in Eqs.\ \eqref{finalgenfnhypoexp_i} and \eqref{finalgenfnhypoexp}, we obtain
\begin{align}
\label{GsHeq}
G_s^H(z)=&\frac{(x-1)}{2xy} \left(\left(\frac{2 x y}{N (x-1)}+1\right)^{N-M} \left(2 \left(\frac{x y}{N (x-1)}+1\right)^{M}-1\right)-1\right) \notag \\ &\times \underset{s=0}{\overset{\infty }{\prod }}\left(1-\frac{2 x y}{x (N+2y)-N 2^s}\right)^{N-M} \left(1-\frac{x y}{x (N+y)-N 2^s}\right)^{M}. 
\end{align}
This can also be written in exponential form as we have previously done for other expressions. The mean and variance of protein fluctuations are given by
\begin{align}
\label{meanmodelIVsp}
\langle n \rangle_s^H&=\frac{\alpha y \left(M(M-1) + 2N(3N-2M+1)\right)}{2 N^2}\notag \\
&=\langle n \rangle_s^E\left(\frac{M+2(N-M)}{N}\right)-\frac{\alpha y M(N-M)}{2N^2}, \\
\label{varmodelIVsp}
\left( \sigma_s^H\right)^2&=\frac{\alpha y \left(M(M-1) + 2N(3N-2M+1) \right)}{2
   N^2} + \frac{\alpha^2 y\left(10N^2 + 2N(3-4M) + 3 M(M-1) \right)}{3 N^2} \notag\\
   &+ \alpha^2 y^2\left(\frac{4 N^4 + 40 N^3 -4 (3 M (M+3)-5) N^2 + 12 M \left(M^2-1\right) N-3 M^2 (M-1)^2 }{12 N^4}\right).
\end{align}
Using the notation $(\sigma_s^E)^2\Bigr|_{vy}$ to denote the variance in Model III, i.e.\ Eq.\ \eqref{varErlang}, when the transcription rate is $vr$ we find 
\begin{align}
    \left( \sigma_s^H\right)^2 &= \left[(\sigma_s^E)^2\Bigr|_y\left(\frac{M}{N}\right) + (\sigma_s^E)^2\Bigr|_{2y} \left(\frac{N-M}{N}\right)\right] \notag \\
    &\qquad - \frac{\alpha y M(N-M)}{2N^2}\left(1+2\alpha -\frac{\alpha y\left((M-1)^2  + N(N-3M-2) \right)}{2N^2} \right).
\end{align}

Note that for the special cases $M = 0$ and $M=N$, these two equations are the same as the mean and variance of Model III (given by Eqs.\ \eqref{meanErlang} and \eqref{varErlang}) with $y$ replaced by $2y$ and $y$, respectively; this is since in this case the transcription rate is the same in all phases of the cell cycle and equal to $2r$ and $r$, respectively. It also follows that Model II is obtained by setting $M=N$ and taking the limit $N \rightarrow \infty$. Hence Model IV contains as special cases the previous Models II and III. Note that while we have considered the cases $M = 0$ and $M=N$ to see the relationships between the various models, when we want to explicitly model replication we need $0<M<N$ and $N\ge 2$ since there is always a pre-replication and post-replication phase of the cell cycle. In Fig.\ 4 we show that the theoretical protein distributions for each phase of the cell cycle accurately match those obtained from stochastic simulations.

The coefficient of variation squared ($CV_s^2=\left(\sigma_s^H/\langle n \rangle_s^H\right)^2$) can be shown to decrease monotonically with increasing $\alpha$ and $y$. However, the dependence on $N$ and $M$ (the replication phase of the cell cycle) is non-monotonic (see Appendix \ref{sec:cv2}). In Fig.\ 5 we show plots of the $CV_s^2$ as a function of all four parameters which numerically verifies the aforementioned properties. We also observe that the size of noise as measured by $CV_s^2$ is almost independent of the replication phase $M$ for large $y$ but increases monotonically with $M$ for small $y$. 

There is an interesting relationship between Model III and Model IV, as follows. If we renormalise the parameter $y$ in Model III by changing it to $\Lambda_s y$ where 
\begin{align}
\label{lambdadefinition}
\Lambda_s=2 -\frac{M}{N} - \frac{M(N-M)}{N(1+3N)},
\end{align}
then $\langle n \rangle_s^H = \langle n \rangle_s^E$ and $\left(\sigma_s^H\right)^2 \approx \left(\sigma_s^E\right)^2$. Note that while the relationship between the means is exact this is not true for the variances; the accuracy of the latter, however, is shown for six different parameter sets in Fig.\ 6a. Note also that since the mean and variance of Model IV and the renormalised Model III match, it follows that if the distribution in both models is well approximated by a negative binomial (a two parameter distribution) then we expect the two distributions to also match. This is indeed the case for models with small $y$ and including sufficient cell cycle length variability, i.e.\ moderate $N$, as can be seen in Fig.\ 6b. 

However, when the cell cycle length variability decreases as $N\to\infty$ we see that the distribution of Model IV starts to deviate from the renormalised Model III when $y$ grows large as we show in Fig.\ 6c. This can be understood from the observation that the pre-replication and post-replication phase of the cell for large $y$ have distinct protein distributions, which was visible in Fig.\ 4 as well, and the fact that the total observation distribution is a sum of both contributions. For moderate $N$, i.e, significant cell cycle length variability, the pre-replication and post-replication protein distributions overlap since they are wide,  which yields good correspondence with Model III as can be seen in Fig.\ 6b (compare solid and dashed blue lines). However for $N$ large enough (low cell cycle length variability) and large $y$, the overlap between the two contributions becomes smaller, resulting in the almost bimodal nature of the distribution in Fig.\ 6c. Hence it follows that replication effects in Model IV can be described well by an appropriately scaled Model III provided cell cycle length variability is not too small. It also follows that it can also be well described by Model I with renormalised parameters since the latter we have shown in Section 4 to be in good agreement with Model III provided $y$ and $N$ are not large.

Finally we mention that similar conclusions hold for population snapshot observations as we have seen for lineage data. A derivation of the snapshot distribution for Model IV can be found in Appendix \ref{sec:popgenfn}. The population equivalent of Eq.\ \eqref{GsHeq} is given by $G^H_p(z) = G_{N,p}(z)W(z)$, where 
\begin{align}
G_{N,p}(z) &= \underset{s=0}{\overset{\infty }{\prod }}\left(1-\frac{2 x y}{x (N 2^{1/N}+2y)-N 2^{1/N} 2^s}\right)^{N-M} \left(1-\frac{x y}{x (N 2^{1/N}+y)-N 2^{1/N} 2^s}\right)^{M},\\
    W(z) &= \left(2^{\frac{1}{N}}-1\right) N (x-1) 
    \notag \\
   &\quad \times \left( \frac{\left(\frac{2 x y}{N (x-1)}+2^{\frac{1}{N}}\right)^{N-M}\left(\left(\frac{x y}{N
   (x-1)}+2^{\frac{1}{N}}\right)^M - 1\right) }{\left(2^{\frac{1}{N}}-1\right) N (x-1)+x y}   +\frac{ \left(\frac{2x y}{N (x-1)}+2^{\frac{1}{N}}\right)^{N-M}-1}{\left(2^{\frac{1}{N}}-1\right) N (x-1)+2 x y}\right).
\end{align}
As shown in Fig.\ 6b, the difference between the population snapshot distribution and the lineage distribution can be significant (compare solid red and blue lines). The mean of the former is less than that of the latter which is due to a preponderance of young cells in population measurements (as discussed in Section 2). Considering the appropriate rescaling of the parameter $y$ by changing it to $\Lambda_p y$, where $\Lambda_p = 2^{1-(M/N)}$, we also get good agreement between Model IV and Model III for population snapshots when the cell cycle length variability is moderate (small $N$)  and $y$ is small (compare dashed and solid red lines in Fig.\ 6b). This however does not carry through for the case of large $y$ and $N$ (Fig.\ 6d). Hence while the distributions are appreciably different for population and lineage cases, nevertheless qualitatively the results for the two are similar. 

\section{Extrinsic noise floor}

Plots of the coefficient of variation squared versus mean protein number have been used in the literature to separate intrinsic noise from extrinsic noise (see in particular Fig.\ 2B of Ref.\ \cite{taniguchi2010quantifying}). Specifically, intrinsic noise is associated with the term proportional to the inverse mean since its contribution decreases with the mean protein number, while extrinsic noise is associated with the term which is independent of the mean. Within this interpretation, using the expressions previously derived for the mean and variance for lineage observations, it is clear that Model I predicts no extrinsic noise ($CV^2$ is inversely proportional to mean) while Models II, III and IV predict extrinsic noise stemming from binomial partitioning and cell cycle length variability. Since Models II and III are special cases of Model IV, we shall only consider the latter. Using Eqs.\ \eqref{meanmodelIVsp}-\eqref{varmodelIVsp} for lineage distributions, we can write the coefficient of variation squared in terms of the mean protein expression level
\begin{equation}\label{eq:noisefloor}
    CV_s^2 = \frac{\mathcal{E}_{s,\text{int}}(M,N,\alpha)}{\langle n\rangle_s^H} + \mathcal{E}_{s,\text{ext}}(M,N),
\end{equation}
where the positive functions $\mathcal{E}_{s,\text{int}}$ and $\mathcal{E}_{s,\text{ext}}$ are given by
\begin{align}\label{eq:noisefloor-decomposed}
    \mathcal{E}_{s,\text{int}}(M,N,\alpha) &= \alpha\left[\frac{6M(M-1)+ 4 N (5 N+3-8M)}{3M(M-1)+ 6 N (3N+1-2M)}\right]+1, \\ \label{magiceq}
    \mathcal{E}_{s,\text{ext}}(M,N) &= \frac{4 N^4+40 N^3 -4 (3 M (M+3)-5) N^2+12 (M^2-1)M N -3 M^2\left(M-1\right)^2}{3 \left(M(M-1) + 2 N (3N+1-2M)\right)^2}.
\end{align}
This shows that in the limit of abundant proteins, the protein noise as measured by the coefficient of variation squared, tends to a constant, $\mathcal{E}_{s,\text{ext}}(N,M)$ which is independent of the intrinsic protein dynamics. This limiting value is only controlled by the cell cycle length variability via $M$ and $N$. The intrinsic noise, on the other hand, is governed by both cell cycle length variability and protein burst size. 

To study the dependencies on cell cycle length variability we let $M=uN$, where $u\in[0,1]$ represents the fraction of the cell cycle spent in the non-replicated phase. It can be shown that if $u$ and $\alpha$ are fixed then the internal component of noise given by the first term in Eq.\ \eqref{eq:noisefloor} increases with $N$ while the external component given by the second term decreases with $N$ (see Appendix C). Hence unexpectedly, $CV_s^2$ can increase or decrease with cell cycle length variability. More specifically in Appendix C (see Eq. \ \eqref{ythreshold}) we show that if $y$ is above a threshold then protein noise increases with increasing cell cycle variability. This condition is met in mammalian cells and yeast cells \cite{schwanhausser2011global,miller2011dynamic} since the range of $y$ and $\alpha$ is greater than one. In contrast in bacteria, due to the very low mean mRNA produced per cell cycle (as low as 0.1 \cite{taniguchi2010quantifying}), the condition is not necessarily met and hence it is possible to have $CV_s^2$ decrease with increasing cell cycle length variability (the dependence is however very weak as can be seen in Fig. \ 7a). Note that one can derive population expressions equivalent to Eqs.\  \eqref{eq:noisefloor-decomposed}-\eqref{magiceq}. It can be proved (see Appendix C) that unlike for lineages, $CV_p^2$ computed from population snapshots always increases with increasing cell cycle length variability (see Fig. \ 7). 

Next we determine bounds on the external noise floor for $N\geq 1$. First of all, we note that since the extrinsic noise floor $\mathcal{E}_{s,\text{ext}}$ decreases monotonically with $N$, its upper bound must be given by $N=1$ while its lower bound by $N \rightarrow \infty$. For the case of purely exponential cell cycle length variability, i.e.\ $N=1$, we then find that
\begin{align}\label{expcaseC1}
    \mathcal{E}_{s,\text{int}}(u,\alpha) &= \left(\frac{4\alpha}{3}+1\right) -  \frac{2 \alpha u(1-u)}{3 (u^2-5u+8)}, \\\label{expcaseC2}
    \mathcal{E}_{s,\text{ext}}(u) &= \frac{1}{3} +\frac{4 u(1-u)(2-u)(4-u) }{3 (u^2-5u+8)^2},
\end{align}
which is maximal when $u\approx 0.55$ at 0.39. On the other hand, in the opposite limit of deterministic cell cycle times, i.e.\ $N\to\infty$, the noise contributions become
\begin{align}
    \mathcal{E}_{s,\text{int}}(u,\alpha) &= \left(\frac{10 \alpha}{9}+1\right) - \frac{8\alpha u (1-u) }{9 (u^2-4u+6)}, \\
    \mathcal{E}_{s,\text{ext}}(u) &= \frac{1}{27} +\frac{4 u(1-u) (7 u^2-22u+12)}{27 (u^2-4u+6)^2},
\end{align}
which is minimal when $u\approx 0.87$ at 0.034. Hence it follows that the external noise floor for $N\geq 1$ is contained in the approximate interval (0.034,0.39). 

Note that for $u=0$ or $u=1$, the special case without a replication phase, Eq.\ \eqref{eq:noisefloor} together with Eqs.\ \eqref{expcaseC1}-\eqref{expcaseC2} almost perfectly recovers a recent result in the literature for the $CV^2$ of protein fluctuations in a model with exponential cell cycle length variability (see Eq.\ 7 of Ref \cite{jkedrak2019exactly}), $CV^2 = (4/3)(\alpha+1/2)/\langle n\rangle + 1/3)$. The same result is obtained for population snapshot calculations. The slight discrepancy is likely caused by the continuum approximation for protein levels in \cite{jkedrak2019exactly}.

An experimental value of approximately 0.1 has been measured for the extrinsic noise floor in the bacterium {\it{E. coli}} (see Fig.\ 2B of \cite{taniguchi2010quantifying}) which falls within the range of our theory (as described above). From the limited lineage data shown in the Taniguchi et al.\ paper (Fig.\ 2C shows three lineages with 9 cell division events), it is not clear which Erlang distribution would best fit their data. However if we consider replication to occur in the middle of the cycle ($M = N/2$) and the cell cycle length to be well described by an Erlang distribution with $N=5$ phases, then Eq.\ \eqref{magiceq} predicts a value of $\mathcal{E}_{s,\text{ext}} \approx 0.12$ which is remarkably close to the experimental value of 0.1. Note that $N=5$ is not unrealistic; it would imply a maximum difference of up to $1/\sqrt{N}\approx40\%$ of the cell cycle length from its mean which is consistent with some recent experiments in {\it{E. coli}} \cite{thomas2017making,hashimoto2016noise}. The noise decomposition above can also be done for population snapshots using the equations for the mean and variance derived in Appendix \ref{sec:popgenfn}. It is found that the theory predicts the same value of $\mathcal{E}_{p,\text{ext}} \approx 0.1$ for extrinsic noise, and hence this result is insensitive to the type of observations made. 

\section{Conclusion}

In this paper we have derived the PGF corresponding to the stable protein number distributions in stochastic gene expression models with cell cycle length variability. Our work has the following special features: (i) the solution method used allows the derivation of distributions rather than the mean and variance; (ii) the distributions for cell cycle length variability assumed by the model are of a very general form (hypoexponential) which fit the majority of experimentally measured distributions; (iii) the model allows the explicit description of the variation of transcription and burstiness with the position of the cell cycle (the cell age); and (iv) the calculations are done for both lineage and population snapshot observations which enhances the match between theory and experiments. A necessary underlying assumption of our approach to compute the PGF is that protein is stable, i.e.\ its decay occurs purely due to dilution by cell division. This is a good assumption when protein lifetimes are much longer than the mean cell cycle time such as in {\it{E. coli}} \cite{maurizi1992proteases} and yeast  \cite{belle2006quantification}. In mammalian cells, about 70 percent of proteins are longer lived than the mean cell cycle time, and hence the approximation is also reasonable \cite{schwanhausser2011global}. 

Special cases of our model can be found in the literature: (i) for a cell cycle composed of $N$ phases, each of which is exponentially distributed in length, and assuming lineage observations, expressions for the mean and variance have been obtained in Ref.\ \cite{soltani2016effects,soltani2016intercellular}; (ii) for a cycle whose length is exponentially distributed, an expression for the approximate protein number distribution (assuming large enough protein numbers) for both lineages and population snapshots has been derived in Ref.\ \cite{jkedrak2019exactly}.

A major contribution of our study is the comparison of different models of gene expression including those with an implicit cell cycle description (Model I) via effective protein degradation and models with an explicit cell cycle description with either regular (Model II) or random interdivision times (Models III and IV). We found that the protein distributions of Models II-IV are well approximated by a negative binomial provided cell cycle length variability is large and $y$ (the mean number of mRNA per cycle) is small (we shall henceforth call these the special conditions). In such cases, the implicit cell cycle model (Model I) with renormalised parameters can describe the results of the explicit cell cycle models. When the special conditions are not met, the distributions show either a flat region near the mode or else have a right shoulder which in some cases can almost look like bimodality; of course, Model I cannot capture these distributions. Such a case may be common for gene expression in mammalian cells and yeast where it is estimated that for many genes, $y$ can take values in the range of 1 to about 600 \cite{schwanhausser2011global,miller2011dynamic}; in contrast in bacteria $y$ has the range $0.1-10$ \cite{taniguchi2010quantifying} and hence deviations from negative binomial distributions are likely much less common. Also we have shown that when the special conditions are not met, the distributions of models including replication or more complex age-dependent transcriptional dynamics cannot be described by models that assume constant transcription through the cell cycle such as those found in \cite{jkedrak2019exactly,thomas2017making}. Our analysis shows that in a model assuming Erlang distributed cell cycle duration and replication time, for lineages, the coefficient of variation squared can either increase or decrease with cell cycle variability whereas for population snapshots, the coefficient of variation squared increases with cell cycle variability. We also show that the the coefficient of variation squared has a complex dependence on the replication time; it is practically independent of the replication time for large $y$ but increases monotonically with the replication time for small $y$. Finally we show that given experimental cell cycle length distributions for {\it{E. coli}} and assuming replication occurs halfway through the cell cycle, our theory predicts a value for extrinsic noise which is within a few percent of that measured in Ref.\ \cite{taniguchi2010quantifying}. 

Despite its generality, our study has a number of limitations: (i) the analytical approach cannot be extended to derive mRNA distributions. This is since mRNA typically degrades faster than the mean cell cycle time \cite{taniguchi2010quantifying} and hence is not stable, which is a necessary assumption to solve for the PGF; (ii) we have assumed binomial partitioning. While this assumption presents the simplest reasonable model of stochastic partitioning, it likely fails for those proteins which are highly localised \cite{huh2011non}; (iii) we have assumed that there is no correlation between the cell cycle duration of mother and daughter cells. Experiments show such a correlation exists \cite{siegal2008tightly,cerulus2016noise}. Overcoming these limitations are key to expanding the realism of the model and are the subject of on-going research. 

\section*{Acknowledgments} 
R.\ P-C.\ acknowledges support from the UCL Mathematics Clifford Fellowship. R.\ G.\ acknowledges support from the Leverhulme Trust (RPG-2018-423). C.\ B.\ acknowledges the Clarendon Fund and New College, Oxford, for funding.

\section*{Appendix} 

\appendix

\section{Derivation of Models I and II from a model with an explicit description of mRNA and of a cell cycle of fixed length}\label{sec:effectivemodels}

\subsection{Derivation of Model I} 

We consider a three-stage gene expression model of the type
\begin{equation}
\label{3stagemodelnoactivedecay}
G \xrightarrow{\sigma_b} G^*,\quad G^* \xrightarrow{\sigma_u} G,\quad G \xrightarrow{\rho_u} G + M,\quad M \xrightarrow{d_m} \emptyset,\quad M \xrightarrow{h} M + P,
\end{equation}
Note that there is no active protein decay and instead we assume protein decay occurs only due to binomial partitioning at cell division. The latter is assumed to happen at regularly spaced time intervals of length $T$. If the rate of promoter switching is very fast compared to the mRNA and protein timescales then there is no need to explicitly model $G^*$. Rather it is sufficient to model expression from a single promoter state with an effective transcription rate equal to the true transcription rate multiplied by the fraction of time that the gene is on. In other words the three-stage model \eqref{3stagemodelnoactivedecay} reduces to the two-stage model
\begin{equation}
\label{2stagemodelnoactivedecay1}
G \xrightarrow{r} G + M,\quad M \xrightarrow{d_m} \emptyset,\quad M \xrightarrow{h} M + P,
\end{equation}
where $r = \rho_u \sigma_u / (\sigma_b + \sigma_u)$ is the effective mRNA transcription rate. Let the PGF be defined as
\begin{equation}
    G(z',z,t)=\sum_{m,n} (z')^m (z)^n P(m,n,t),
\end{equation}    
where $P(m,n,t)$ is the probability of observing $m$ mRNAs and $n$ proteins at time $t$. The PGF then satisfies a PDE which when non-dimensionalised on the cell cycle time scale, $t=T\tau$, is given by
\begin{equation}\label{eq:twostagePGF}
\frac{\partial G}{\partial \tau} = r T(z'-1)G+ d_m T(1-z')\frac{\partial G}{\partial z'} + h T z'(z-1)\frac{\partial G}{\partial z'}.
\end{equation}
Furthermore binomial partitioning under cyclo-stationarity conditions \cite{berg1978model} leads to the boundary condition (in non-dimensional form) 
\begin{equation}\label{eq:twostageBC}
    G(z',z,0) = G\left(\frac{z'+1}{2},\frac{z+1}{2},1\right).
\end{equation}
By using the definitions of the mean numbers of proteins and mRNA 
\begin{align}
\langle n \rangle (\tau) = \frac{\partial G(z',z,\tau)}{\partial z}\bigg|_{z'=z=1}, \langle m \rangle (\tau) = \frac{\partial G(z',z,\tau)}{\partial z'}\bigg|_{z'=z=1},    \end{align}
it is straightforward to show that the time-evolution of the means is given by the coupled ODEs
\begin{align}
\frac{\mathrm{d} \langle n \rangle (\tau)}{\mathrm{d} \tau} &= hT  \langle m \rangle (\tau), \\  
\frac{\mathrm{d} \langle m \rangle (\tau)}{\mathrm{d} \tau} &= rT -d_mT\langle m \rangle (\tau),  
\end{align}
with the boundary conditions $2\langle n \rangle (0) = \langle n \rangle (1)$ and $ 2\langle m \rangle (0) = \langle m \rangle (1)$.
Solving these ODEs one obtains
\begin{equation}
\langle n \rangle (\tau) = \frac{h r e^{-d_m T \tau} \left(e^{\tau  T d_m} \left(1-(\tau +1) T
   d_m\right)+2 e^{(\tau +1) T d_m} \left((\tau +1) T d_m-1\right)+e^{T
   d_m}\right)}{d_m^2 \left(2 e^{T d_m}-1\right)},   
\end{equation}
where $\tau$ is to be understood as the fractional cell age which equals zero when a cell is born and one just before a cell divides. Note that we do not show the equation for the mRNA since it is not relevant to our analysis. Hence it follows that the mean number of proteins in single lineage and population measurements is given by
\begin{align}
  \langle n \rangle_s &= \int_{0}^1 \langle n \rangle (\tau) \mathrm{d}\tau = \frac{\alpha  y \left(d_m \left(3 T d_m-2\right)+\frac{1}{T-2 T e^{T
   d_m}}+\frac{1}{T}\right)}{2 T d_m^2}, \\
   \langle n \rangle_p &= \log(2) \int_{0}^1 2^{1 - \tau} \langle n \rangle (\tau) \mathrm{d}\tau = \frac{\alpha  T y d_m}{T \log (2) d_m+\log ^2(2)},
\end{align}
where we used $y=rT$ and $\alpha = h/d_m$. Note that here we used the fact that when interdivision times are regularly spaced in time, the probability of observing a cell of age $t \in[0,T]$ is uniform for single lineage measurements and equal to $2^{1-t/T}\log2/T$ for population measurements \cite{berg1978model} (see also Appendix D for a derivation of the latter). Taking the limit that mRNA decays fast compared to the cell cycle length, i.e.\ $d_m T \rightarrow \infty$, we obtain
\begin{align}
\label{nmeansingle}
    \langle n \rangle_s &\approx \frac{3 \alpha  y}{2}, \\
    \label{nmeanpop}
    \langle n \rangle_p &\approx \frac{\alpha  y}{\log (2)}.
\end{align}
Now one may ask what constitutes an effective system of reactions that describes protein dynamics and which has the same mean number of proteins as above. If we consider this effective system to be given by Model I, that is the set of reactions
\begin{equation}
G \xrightarrow{r} G + m P,\quad P \xrightarrow{d'} \emptyset,
\end{equation}
where $m$ is a geometrically distributed random number with mean $\alpha$, it follows that the mean number of proteins is given by
\begin{align}
\label{effectivemeanI}
 \langle n \rangle_{I} = \frac{r \alpha}{d'} = \frac{\alpha y}{d'T}.
\end{align}
Equating Eq.\ \eqref{effectivemeanI} with Eq.\ \eqref{nmeansingle}, we obtain the effective protein degradation rates for single lineage measurements,
\begin{align}
d'= \frac{2}{3T}.
\end{align}
Similarly equating Eq.\ \eqref{effectivemeanI} with Eq.\ \eqref{nmeanpop}, we find that the effective protein degradation rates for population measurements is
\begin{align}
d'= \frac{\log(2)}{T}.
\end{align}

\subsection{Derivation of Model II}

Starting from the same mRNA model as used in A.1 we now derive Model II. Consider the PDE Eq.\ \eqref{eq:twostagePGF} with boundary condition Eq.\ \eqref{eq:twostageBC}. To solve this PDE, we employ the method of characteristics which yields the system of equations
\begin{align}
\frac{\mathrm{d} \tau}{\mathrm{d}s} &= 1,\\
\frac{\mathrm{d} z'}{\mathrm{d}s} &= d_m T(z'-1) + h T z'(1-z),\\
\frac{\mathrm{d} z}{\mathrm{d}s} &= 0,\\
\frac{\mathrm{d} G}{\mathrm{d}s} &= r T(z'-1) G,
\end{align}
where $s$ is the characteristic variable. In particular we see that $\tau=s$ and $z=z_0$ is constant. 

To consider the case of unstable mRNA we define $\varepsilon=1/(d_m T)$ as the asymptotic variable.  We see that to achieve a dominant balance as $\varepsilon\to 0$ (or $d_m T\to \infty$) we could choose $h T=\mathcal{O}(d_m T)$.
As before we use, $\alpha = \lim_{d_m T\to\infty} (h T/d_m T)$ and taking this scaling we see that as $\varepsilon\to 0$ we get
\begin{equation}
\varepsilon \frac{\mathrm{d} z' }{\mathrm{d}s} = z'-1 + \alpha z'(1-z) + \mathcal{O}\left(\varepsilon\right),
\end{equation}
and therefore have to leading order $(z'-1) - \alpha z'(z-1)=0$. This results in $z'\to 1/(1-\alpha(z-1))$ quickly in the limit of $\varepsilon\to 0$ and therefore the following effective PDE for the PGF 
\begin{equation}
\frac{\partial G}{\partial \tau} = r T\frac{\alpha(z-1)}{1-\alpha(z-1)}G.
\end{equation}
Note that we can then define $\tilde{G}(z,\tau) = \lim_{z'\to 1}G(z',z,\tau)= \sum_n y^n P(n,\tau)$, i.e.\ the PGF for the marginal protein distribution. It follows now that $\tilde{G}$ satisfies the following cyclo-stationary system in the limit of $\varepsilon\to 0$
\begin{subequations}\label{eq:modelII}
\begin{align}
\frac{\partial \tilde{G}}{\partial \tau} &= r T\frac{\alpha(z-1)}{1-\alpha(z-1)}\tilde{G}, \\
\tilde{G}(z,0) & = \tilde{G}\left(\frac{z+1}{2},1\right).
\end{align}
\end{subequations}
Note that this is the same system satisfied by $G(z,t)$ from Model II, Eqs.\ \eqref{PGFpdeModelII} and \eqref{partition1} in the cyclo-stationary limit where we drop the superscript for the generation. Therefore we have shown that the two-stage expression model \eqref{2stagemodelnoactivedecay1} with binomial partitioning converges to model II in the limit of fast mRNA decay compared to the cell cycle length, i.e.\ $d_m T\gg 1$.  

\section{Stochastic simulations} 
Stochastic simulations in this work were carried out using a method similar to the First Division Algorithm in \cite{thomas2017making}, which effectively is the (modified) next reaction method with the addition of cell division and observation events. Below follows a description of the exact procedures used to calculate stochastic realisations of Models II, III and IV in this paper. The code used for this work can be found online at \url{https://bitbucket.org/CasperBeentjes/protein-cell-cycle-ssa}.
\subsection*{Lineage simulation}
For the simulation of lineage data we consider a single cell in a mother machine that we continuously track. Measurements of the cell's contents are taken at intervals defined by some observation time distribution. For this paper we take measurements at regular intervals, i.e.\ a delta distribution for the observation interval distribution.
\begin{itemize}
    \item [1.] \textit{Initialisation.} Start a cell in the mother machine at time $t=0$ with initial molecule content $n$. Assign a phase $j$ and age within the phase, $a$.
    \item [2.] Generate waiting times $\tau_r$, $\tau_p$ and $\tau_o$, the time until the next biochemical reaction, the next phase change and the next observation respectively.
    \item [3.] Until $t\geq t_{\text{final}}$. Pick $\Delta=\min(\tau_r,\tau_p,\tau_o)$, update all waiting times ($\tau_r,\tau_p,\tau_o$) by $\tau\to\tau-\Delta$ and update phase age $a\to a+\Delta$. Based on the minimum found for the waiting times proceed to:
    \begin{itemize}
        \item [a.] \textit{Biochemical reaction.} Select the biochemical reaction occurring, e.g.\ using Gillespie-type algorithm, and update cell molecule content. Generate a new time until next biochemical reaction $\tau_r$.
        \item [b.] \textit{Phase progression.} Based on the current phase of the cell proceed to:
        \begin{itemize}
            \item [i.] If phase of the cell is $N$, reset the new phase of the cell to the first phase, $j\to 1$. Binomially partition the cell contents across the cell in mother machine and daughter cell, discard daughter cell and keep following cell in mother machine.
            \item [ii.] Otherwise, set the new phase of the cell, $j\to j+1$.
        \end{itemize}
        Set the age of cell in the new phase to $a=0$ and update the biochemical rate parameters according to the current phase $j$. Generate a new time until next phase progression $\tau_p$ and a new time until next biochemical reaction $\tau_r$.
        \item [c.] \textit{Observation.} Write the current cell contents and other quantities of interest to disk. Generate a new time until next observation $\tau_o$.
    \end{itemize}
    Update simulation time $t \to t+\Delta$ and return to the start of step 3.
\end{itemize}

\subsection*{Population snapshot simulation}

For the simulation of population snapshot data we consider an initial batch of cells that we let grow and divide until some final time $t_{\text{final}}$ at which point we measure the contents of all the cells present. 
\begin{itemize}
    \item [1.] \textit{Initialisation.} Start batch of cells at time $t_i=0$ with initial molecule contents $n_i$, where $i$ index rolls over the batch of present cells. Assign a phase $j_i$ and age within the phase, $a_i$, to each cell in the batch. Start with the first cell in the batch, $l\to 1$.
    \item [2.] Until all cells in batch have reached final time $t_{\text{final}}$ pick the next cell in the batch and proceed to step 3.
    \item [3.] Generate waiting times $\tau_r$ and $\tau_p$ the time until the next biochemical reaction and the next phase change respectively. Set the time until observation as $\tau_o\to t_{\text{final}}-t_l$.
    \item [4.] Until $t_i= t_{\text{final}}$. Pick $\Delta=\min(\tau_r,\tau_p,\tau_o)$, update all waiting times ($\tau_r,\tau_p,\tau_o$) by $\tau\to\tau-\Delta$ and update phase age $a\to a+\Delta$. Based on the minimum found for the waiting times proceed to:
    \begin{itemize}
        \item [a.] \textit{Biochemical reaction.} Select biochemical reaction occurring, e.g.\ using Gillespie-type algorithm, and update cell molecule content. Generate a new time until next biochemical reaction $\tau_r$.
        \item [b.] \textit{Phase progression.} Based on the current phase of the cell proceed to:
        \begin{itemize}
            \item [i.] If phase of the cell is $N$, reset the new phase of the cell to the first phase, $j\to 1$. Binomially partition the cell contents across the current cell and a daughter cell. Add the daughter cell with its content to the batch of cells with start time $t_{i'}=t_l$, $a_{i'}=0$ and $j_{i'}=1$.
            \item [ii.] Otherwise, set the new phase of the cell, $j\to j+1$.
        \end{itemize}
        Set the age of cell in the new phase to $a=0$ and update the biochemical rate parameters according to the current phase $j$. Generate a new time until next phase progression $\tau_p$ and a new time until next biochemical reaction $\tau_r$.
        \item [c.] \textit{Observation.} Write the current cell contents and other quantities of interest to disk. Let $l\to l+1$ and return to step 2.
    \end{itemize}
    Update simulation time $t_l \to t_l+\Delta$ and return to the start of step 4. 
\end{itemize}

\section{Coefficient of variation proofs for Model IV}\label{sec:cv2}
\subsubsection*{Effect of burst size $\alpha$ and mRNA production per cycle $y$}
First we prove that $CV^2$ (and $CV$) monotonically decrease with increasing $\alpha$ and $y$. We note using Eqs.\ \eqref{meanmodelIVsp}-\eqref{varmodelIVsp} that the mean and variance of protein numbers in Model IV with replication can be written as
\begin{align*}
\left(\langle n \rangle_s^H\right)^2&=\alpha^2 y^2 A, \\
\left( \sigma_s^H\right)^2&=\alpha y B + \alpha^2 y C + \alpha^2 y^2 D,
\end{align*}
where $A,B,C,D$ are functions of $M,N$ solely. First we observe that $A> 0$ must hold by the observation that the mean of the Model IV is necessarily positive if $\alpha>0$ and $y>0$. This means that we have for the squared coefficient of variation
\begin{equation*}
    CV^2 = \frac{1}{A}\left(\frac{B}{\alpha y} + \frac{C}{y} + D \right).
\end{equation*}
Note that $CV^2\geq 0$ must hold for all $\alpha\geq 0$ and $y\geq 0$. By considering the limit $\alpha\downarrow 0$ we deduce that $B\geq 0$. Similarly, by considering the limit $y\downarrow 0$ we find that for all $\alpha\geq 0$ we must have $B + C\alpha\geq 0$ and therefore $C\geq 0$. Finally from the limit $y\to \infty$ we deduce that $D\geq 0$. It then immediately follows that $CV^2$ and $CV$ decrease monotonically with $\alpha$ and/or $y$ increasing. Note that this argument applies equally well to the population snapshot case.

\subsubsection*{Effect of cell cycle variability $N$ for lineages}

Next we show that the extrinsic noise floor, i.e.\ the limit of $CV^2$ for abundant proteins, is a monotonic decreasing function of $N$. Recalling Eq.\ \eqref{magiceq} and re-writing the noise floor $\mathcal{E}_{s,\text{ext}}(M,N)$ using $M=uN$, where $u\in[0,1]$, we have
\begin{equation}
    \mathcal{E}_{s,\text{ext}}(u,N) = \frac{N^2 \left(-3 u^4+12 u^3-12 u^2+4\right)+N \left(6 u^3-36 u+40\right)-3 u^2-12 u+20}{3 \left(N \left(u^2-4
   u+6\right)-u+2\right)^2}.
\end{equation}
We then note that the derivative of the noise floor with respect to $N$ is given by
\begin{equation}
    \frac{\mathrm{d} \mathcal{E}_{s,\text{ext}}}{\mathrm{d} N} = -\frac{4 \left((-3 u^3+16 u^2-48 u+40) + N(3 u^4-18 u^3+58 u^2-92 u+56)\right)}{3 (N (u^2-4u + 6)+(2-u))^3}<0.
\end{equation}
The inequality follows from considering every term in brackets for $N>0$ and $u\in[0,1]$ and showing it is positive. As a result, we see that the external noise floor is a monotonic decreasing function of increasing $N$, which can be interpreted as decreasing cell cycle length variability.

Next we show that $CV_s^2$ is not monotonic in $N$ for single cell lineage measurements. Recalling Eqs.\ \eqref{meanmodelIVsp} and \eqref{eq:noisefloor-decomposed}, we can re-write the internal noise component and protein mean using $u$ and $N$ as 
\begin{align}
    \mathcal{E}_{s,\text{int}}(u,N) &=1 + 2\alpha - \frac{8 \alpha  N (2-u)}{3 N (6 - 4u + u^2)+3(2-u)},\\
    \langle n\rangle^H_s &= \left(\frac{6-4u+u^2}{2} + \frac{2-u}{2N} \right)\alpha y.
\end{align}
Note that it then follows that 
\begin{align}
    \frac{\mathrm{d} \mathcal{E}_{s,\text{int}}}{\mathrm{d} N} &= -\left(\frac{8   (2-u)^2}{3 \left(N \left(u^2-4 u+6\right)+(2-u)\right)^2}\right)\alpha<0,\\
    \frac{\mathrm{d} \langle n\rangle^H_s    }{\mathrm{d} N} &= - \left(\frac{2-u}{2N^2}\right)\alpha y<0,
\end{align}
which does not allow us to immediately make a conclusion about the behaviour of the ratio of the two quantities. Therefore we note that the internal component of the coefficient of variation squared in fact grows as $N$ increases as shown by
\begin{equation}
    \frac{\mathrm{d} (\mathcal{E}_{s,\text{int}}/\langle n\rangle^H_s  )}{\mathrm{d} N} = \frac{2 (2-u) \left(2 \alpha  \left(N \left(3 u^2-4 u+2\right)+3(2-u)\right)+3 N
   \left(u^2-4 u+6\right)+3(2-u)\right)}{3 \alpha  y \left(N \left(u^2-4
   u+6\right)+(2-u)\right)^3}>0.
\end{equation}
Since the external component of the coefficient of variation squared has earlier been proved to decrease with $N$, it follows that for some positive functions $a_1,a_2$ of $u,N,\alpha$ we find
\begin{equation}
    \frac{\mathrm{d} CV_s^2}{\mathrm{d}N} = \frac{a_1}{y} -a_2,
\end{equation}
which shows that $CV_s^2$ can be both increasing or decreasing as a function of $N$, depending on the exact values of $\alpha, y$. In fact, by considering $a_1$ and $a_2$ and their extreme behaviour (which is for $u=1$) one can show that for 
\begin{equation}
\label{ythreshold}
    y>\frac{3+N}{5+7N} + \frac{1}{\alpha}\frac{3+9N}{2(5+7N)},
\end{equation}
the protein noise for lineages is monotonically decreasing, whereas when this condition is not satisfied we can find $u=M/N\in[0,1]$ so that, perhaps counter-intuitively, decreasing cell cycle variability, i.e.\ increasing $N$, leads to an increase in protein noise. On the other hand this shows that if mRNA production is large enough the protein noise does become monotonic as a function of cell cycle variability.

\subsubsection*{Effect of cell cycle variability $N$ for population snapshots}

For population snapshots we note from results in Appendix D, Eqs.\ \eqref{eq:meanIVpop} and \eqref{eq:varIVpop} in particular, that the internal noise component is composed of
\begin{align}
    \mathcal{E}_{p,\text{int}}(u,N) &=1 + 2\alpha -\frac{\alpha (2-u) 2^{1+u}}{3}\frac{N \left(2^{1/N}-1\right)}{ 2^{1/N}} ,\\
    \langle n\rangle^H_p &=  \frac{2^{1-u}}{N(2^{1/N}-1)}\alpha y.
\end{align}
From here it is easy to show that $\langle n\rangle^H_p$ and $\mathcal{E}_{p,\text{int}}(u,N)$ monotonically grow and decay respectively as $N$ increases. As a result we note that their ratio, the internal noise component $(\mathcal{E}_{p,\text{int}}/\langle n\rangle^H_p  )$, decays monotonically when $N$ increases, i.e.\ the opposite scenario to what happens for single cell lineage measurements.

Finally we will show that $\mathcal{E}_{p,\text{ext}}$ decays monotonically when $N$ increases and as a result that, in contrast to the lineage framework, the total protein noise $CV^2_p$ is always decreasing as a function of increasing $N$, regardless of the values of $\alpha,y$. We note that this proof is more cumbersome than for the lineage measurements and we only sketch the details here. We start by noting that, with $u\in[0,1]$ as before,
\begin{equation}
\begin{split}
    \mathcal{E}_{p,\text{ext}}(u,N)&=-\frac{2^{u}}{6} \frac{N \left(2^{1/N}-1\right) }{2^{2/N}} \left(2^u \left(3 \left(2^{\frac{1}{N}}-1\right) N
   (u-2)^2+2^{\frac{1}{N}} (16-9 u)+3 u-4\right)-12\cdot 2^{1/N}\right)\\
   &\qquad +\left(-4^u + 3\cdot 2^u -1\right) 
   \end{split}
\end{equation}
To make computations slightly more tractable we then make the transformation to $q=2^{1/N}$ (which yields $q> 1$ for $N>0$ and $q\to 1$ when $N\to\infty$) and note that $\mathrm{d}q/\mathrm{d}N<0$ for all $N$. This leaves us to show that $\mathrm{d}\mathcal{E}_{p,\text{ext}}/\mathrm{d}q> 0$ for all $q> 1$ in order to show that the external noise component satisfies $\mathrm{d}\mathcal{E}_{p,\text{ext}}/\mathrm{d}N < 0$. We start with
\begin{equation}
    h_1(q) = \frac{\mathrm{d}\mathcal{E}_{p,\text{ext}}}{\mathrm{d}q}=\underbrace{\frac{2^{u}\log 2}{6 (q \log q)^3}}_{> 0} h_2(q).
\end{equation}
From here we note that $\lim_{q\downarrow 1}h_2(q)=\lim_{q\downarrow 1}h_2'(q)=\lim_{q\downarrow 1}h_2''(q)=0$ and
\begin{equation}
    \frac{\mathrm{d}^3h_2}{\mathrm{d}q^3} = \underbrace{\frac{1}{q^3}}_{> 0}h_3(q).
\end{equation}
The new function $h_3$ satisfies for $u\in[0,1]$
\begin{align*}
    \lim_{q\downarrow 1}h_3(q) &= 6 \left(2^u \left(u^2 \log (8)-12 u \log (2)+2+\log (4096)\right)-6\right) > 0\\
    \frac{\mathrm{d} h_3}{\mathrm{d}q} &= \underbrace{\frac{1}{q}}_{> 0}h_4(q).
\end{align*}
In a similar fashion we proceed to show that for $h_4$ and $u\in[0,1]$ we can write
\begin{align*}
    \lim_{q\downarrow 1}h_4(q) &= 2 \left(2^u (u ((u-4) \log (8)-48)+78+\log (4096))-42\right)>0\\
    \lim_{q\downarrow 1}h_4'(q) &= 2 \left(3\cdot 2^u \left(u^2 \log (2)-2 u (11+\log (4))+38+\log (16)\right)-78\right)>0\\
    \frac{\mathrm{d}^2 h_4}{\mathrm{d}q^2} &= \underbrace{\frac{8}{q}}_{> 0}h_5(q),
\end{align*}
where $h_5(q)=q \left(2^u (16-9 u)-12\right)+\left(-3\cdot 2^u u+5\cdot 2^u-3\right)$ is a positive linear function in $q$ when $u\in[0,1]$. By  arriving at $h_5>0$ for $q> 1$ we find cascading back that $h_4'',h_4',h_4>0$ for $q> 1$. This then proves $h_3',h_3>0$ for $q> 1$ which in turn shows that $h_2''',h_2'',h_2',h_2>0$ for $q> 1$. Finally having proven that $h_2>0$ we find $\mathrm{d}\mathcal{E}_{p,\text{ext}}/\mathrm{d}q=h_1>0$ for $q> 1$, which was what we needed to prove to show that $\mathrm{d}\mathcal{E}_{p,\text{ext}}/\mathrm{d}N<0$.

% \begin{equation}
% \begin{split}
%     h_2(q) &= 2 \left(2^u (2 q (3 u-5)-3 u+4)+6 q\right) \log ^2(q)+(q-1)^2 2^{u+1} (u-2)^2 \log (8) \\
%     &\quad -(q-1) \log (q) \left(2^u (q (9 u-16)+u (u  \log (64)-3-8 \log (8))+4+8 \log (8))+12 q\right)
%   \end{split}
% \end{equation}

% \begin{equation}
% \begin{split}
%     h_3(q) &= 2^u \left(\log (64) \left(q (u-2)^2+2 u^2+8\right)-2 (q (q (9 u-16)+6 u-10)-15 u+8 u \log (8)+20)\right)\\
%     &\quad +2^u\left(-8 (3 (q+1) u-5 q-4) \log(q)\right)-12 q (2 q+2 \log (q)+1)
%   \end{split}
% \end{equation}

% \begin{equation}
% \begin{split}
%     h_4(q) &= 2 \left(2^u (q (q (32-18 u)+u ((u-4) \log (8)-18)+30+\log (4096))\right)\\
%     &\quad +2\left(4 q (5-3 u) \log (q)-12 u+16)-6 q (4 q+2 \log (q)+3)\right)
% \end{split}
% \end{equation}

\section{Population snapshots}\label{sec:popgenfn}

In this Appendix, we correct for the fact that in a population observation, the cell phases and observation times differ from those in single-cell measurements. In order to construct the PGF for the population observations we start from the expression
\begin{equation}\label{eq:PGF-pop}
G_p(z) = \sum_{i=1}^N \Pi_{i,p} G_{i,p}(z),
\end{equation}
where $G_{i,p}(z)$ denotes the PGF for phase $i$ in the population measurement case and $\Pi_{i,p}$ is the probability to observe a cell in phase $i$. Note that we take the distributions for each phase to be normalised to unity to reflect that the $G_{i,p}$ are in fact the marginal distributions for protein in each phase $i$. We will next show how to derive expressions for $G_{i}(z)$ and $\Pi_{i,p}$.

\subsubsection*{Phase distribution}

While the probability of observing a cell in cell cycle phase $i$, denoted as $\Pi_i$, was obvious when performing single-cell measurements, this is not the case for population measurements. This is since each time a cell divides, two cells start in phase $1$ and hence we generally expect that the probability of being in phase $i$ decreases with $i$. We next derive an expression for the probability $\Pi_{i,p}$ for population measurements. 

Let the average number of cells under cyclo-stationary assumptions in cell cycle phase $i$ at time $t$ be denoted by $\langle C_i(t)\rangle$. Then it immediately follows from the model specification that when each phase of the cell cycle is exponentially distributed with parameter $k$, the time-evolution equations are given by
\begin{align}
\frac{\mathrm{d} \langle C_1(t) \rangle}{\mathrm{d} t} &= -k \langle C_1(t) \rangle + 2k \langle C_N(t) \rangle \label{eq.ME1}, \\
\frac{\mathrm{d} \langle C_i(t) \rangle}{\mathrm{d} t} &= -k \langle C_i(t) \rangle + k \langle C_{i-1}(t) \rangle, \quad i=2,\dots,N \label{eq.ME2}.
\end{align} 
The factor 2 in the first equation stands for cellular division: every time a cell divides (leaving phase $N$), two cells start in phase 1. In the case that we are tracking single cell lineages then the factor 2 is replaced by 1. More generally, for cases with asymmetric division (after division some cells differentiate) this factor 2 can be replaced by a factor $\nu\in[0,2]$. We are interested in finding the probability that a cell is in phase $i$, which is given by
\begin{equation}
\label{pidef}
\Pi_i(t) = \frac{\langle C_i(t)\rangle}{\sum_{i=1}^N\langle C_i(t)\rangle}.
\end{equation}
Note that this expression is valid for both population snapshot and lineage observations. While this quantity will change with time, it will eventually approach a steady-state value and this is what we are here interested in. We make the ansatz that given long enough time, $\langle C_i(t)\rangle = \lambda_i \langle C_1(t)\rangle$ where $\lambda_i$ are some time-independent constants which need to be determined (except for $\lambda_1 = 1$ which follows immediately). Substituting this assumption in Eq.\ (\ref{pidef}) we find
\begin{equation}
\label{pidef1}
\Pi_{i,p} = \frac{\lambda_i}{1 + \sum_{j=2}^N \lambda_j}.
\end{equation}
Hence next we determine the values of $\lambda_i$. Substituting the ansatz in Eqs.\ (\ref{eq.ME1})-(\ref{eq.ME2}), we find the relationship:
\begin{equation}
\frac{\lambda_{i-1}}{\lambda_i} = 2 \lambda_N, \quad i=2,\dots,N,
\end{equation}
with $\lambda_1 = 1$. Solving this recurrence relation one finds
\begin{align}
\lambda_i = 2^{(1-i)/N}. 
\end{align}
Substituting the latter expression in Eq.\ (\ref{pidef1}) and simplifying one finally obtains
\begin{equation}
\label{pidef2}
\Pi_{i,p} = \frac{2^{1/N}-1}{2^{(i/N)-1}}, \quad i=1,\dots,N,
\end{equation}
which in the limit of $N\to\infty$ yields the familiar age structure in a population which doubles at mitosis, i.e.\ $f(t)=2^{1-t/T}\log 2 /T$. Note that this result is in contrast to the phase distribution in the single cell measurement case, which was given by a uniform distribution, i.e.\ $\Pi_{i,s}=1/N$ for all $i$. In a population we are more likely to observe cells in an early cell phase compared to lineage data. Note that incidentally we can derive the population growth rate (see next Section) from this formalism, which is necessary to derive the age distribution in a population for each phase.

\subsubsection*{Population growth rate}

To derive the population growth rate, we consider $\langle C(t) \rangle = \sum_i \langle C_i(t)\rangle$, the expected total number of cells, i.e.\ the size of the population. From Eqs.\ \eqref{eq.ME1} and \eqref{eq.ME2} we then find in the long-time limit when the proportion of cells in each phase relative to the total number of cells, i.e.\ $\Pi_{i,p}$, becomes stable that the growth of the cell population is given by
\begin{equation}
    \frac{\mathrm{d} \langle C(t) \rangle}{\mathrm{d} t} = k\langle C_N(t) \rangle = k \Pi_{N,p} \langle C(t) \rangle = k(2^{1/N}-1)\langle C(t) \rangle.
\end{equation}
This shows that the average number of cells in the population grows like $\langle C(t) \rangle \propto e^{\lambda t}$ where $\lambda=k(2^{1/N}-1)$. Note that in the limit of $N\to\infty$ and $k=N/T$ this becomes $\lambda = \log 2/T$, showing that the population doubles in size after every cell cycle when we consider the case of a deterministic cell cycle of length $T$.

\subsubsection*{Population age distribution}

Let's consider $\langle C_i(t,\tau) \rangle$ the average number of cells in a population that are in cell phase $i$ at time $t$ that have been in that cell phase for a duration $\tau$, i.e.\ are of an age $\tau$. After a small time duration $\delta$, all the cells will either advance to an age $\tau+\delta$ or advance to the next cell phase. Therefore we can write the conservation equation
\begin{equation}
\langle C _i(t+\delta,\tau+\delta)\rangle = \langle C_i(t,\tau)\rangle - \langle C_i(t,\tau) \rangle k \delta, \label{eq.conservation}
\end{equation}
where $k$ is the rate of advancing to the next cell phase. Assuming that there is a stationary distribution for the age of the cell population at a phase $i$, $\pi_i(\tau)$, we can write $\langle C_i(t,\tau) \rangle$ as $\langle C_i(t,\tau) \rangle = \langle C_i(t)\rangle \pi_i(\tau)$, where $\langle C_i(t)\rangle$ as before is the expected number of cells in cell phase $i$ at time $t$. Introducing this factorisation of $\langle C_i(t,\tau)\rangle $ in Eq.\ \eqref{eq.conservation}, and taking the limit $\delta\rightarrow 0$, we get the differential equation

\begin{equation}
\frac{\mathrm d \langle C_i(t)\rangle}{\mathrm d t}\pi_i(\tau)+\frac{\mathrm d \pi_i(\tau)}{\mathrm d \tau} \langle C_i(\tau)\rangle = -\langle C_i(t) \rangle \pi_i(\tau)k, \label{eq.diff}
\end{equation}
where we have used the chain rule to compute the derivative of $\mathrm d \langle C_i(x,x)\rangle /\mathrm d x|_{t,\tau}$. Given that in the long time limit, the population of cells is growing at an average rate $\lambda$, which we derived in the previous section, and since the probability of finding a cell in a certain phase $i$, $\Pi_i$, is constant in time, the number of cells in a given phase has to grow with the same rate as the population. This implies that therefore $\mathrm d\langle C_i(t)\rangle/\mathrm d t=\lambda \langle C_i(t)\rangle $. Introducing this equality in Eq.\ \eqref{eq.diff}, we get an equation solely for $\pi_i(\tau)$

\begin{equation}
\lambda \pi_i(\tau) +\frac{\mathrm d \pi_i(\tau)}{\mathrm d \tau} = -\pi_i(\tau)k,
\end{equation}
which can be solved by
\begin{equation}
\pi_i(\tau) = \left(\lambda+k\right)\mathrm e^{-(\lambda+k)\tau}= k2^{1/N}\mathrm e^{-k2^{1/N}\tau}.  \label{eq.p}
\end{equation}

Finally we note that there is a direct link between the age distribution at measurement and the phase length distribution on a population level. In order to achieve a stable distribution for the age at measurement in the long-time limit we note that the population phase length distribution has to be consistent with the age distribution. Since the population age distribution at observation in each phase is still exponential, but now with rate $k2^{1/N}$, the distribution for the time it takes to progress to the next cell phase in a population must also be exponentially distributed with the same rate. This means that in a population of cells we observe a quicker progression through the phases (in distribution on a population level) than when following a single cell. This can be understood from the observation that cells that quickly progress through their phases (and thus divide quickly) will be more abundant in populations than in single cell lineages.

\subsection*{Population snapshot observation distribution}

By the results of the previous section, we can derive the distribution for population snapshot measurements from the single cell measurement statistics via a rescaling of the phase progression rates. The distribution for the length of the cell phases is still exponential which means the Markovian nature of the problem remains intact in the population case. This implies that in all the equations for the PGF derived earlier (for single lineages) we replace $k\mapsto k2^{1/N}$ to find the corresponding population equations. The recipe for calculating population observations therefore is
\begin{itemize}
\item[1] To obtain $G_{i,p}(z)$ replace $k\mapsto k 2^{1/N}$ in the expressions for the lineage distributions $G_{i,s}(z)$, e.g.\ in Eq.\ \eqref{eqGrecH}.
\item[2] Average over the different phases via Eq.\ \eqref{eq:PGF-pop} using $\Pi_{i,p}$ from Eq.\ \eqref{pidef2}.
\end{itemize}

Using these steps we can then solve for the snapshot observation distribution for Model IV with replication which we will write as $G_p(z) = G_{N,p}(z)W(z)$, where 
\begin{align}
\label{genfnNfinalpop}
G_{N,p}(z) &= \underset{s=0}{\overset{\infty }{\prod }}\left(1-\frac{2 x y}{x (N 2^{1/N}+2y)-N 2^{1/N} 2^s}\right)^{N-M} \left(1-\frac{x y}{x (N 2^{1/N}+y)-N 2^{1/N} 2^s}\right)^{M},\\
\label{genfnWfinalpop}
    W(z) &= \left(2^{\frac{1}{N}}-1\right) N (x-1) 
    \notag \\
   &\quad \times \left( \frac{\left(\frac{2 x y}{N (x-1)}+2^{\frac{1}{N}}\right)^{N-M}\left(\left(\frac{x y}{N
   (x-1)}+2^{\frac{1}{N}}\right)^M - 1\right) }{\left(2^{\frac{1}{N}}-1\right) N (x-1)+x y}   +\frac{ \left(\frac{2x y}{N (x-1)}+2^{\frac{1}{N}}\right)^{N-M}-1}{\left(2^{\frac{1}{N}}-1\right) N (x-1)+2 x y}\right).
\end{align}
This result then yields the explicit snapshot distribution for all the other models considered in this paper. We recover the special case of Model III by taking $M=N$ which in turn simplifies the observation distribution to 
\begin{align}
\label{genfnNfinalIIIpop}
G_{N,p}(z) &= \underset{s=0}{\overset{\infty }{\prod }}\left(1-\frac{x y}{x (N 2^{1/N}+y)-N 2^{1/N} 2^s}\right)^{N},\\
\label{genfnWfinalIIIpop}
    W(z) &= \left(2^{\frac{1}{N}}-1\right) N (x-1)  \frac{\left( \frac{x y}{N
   (x-1)}+2^{\frac{1}{N}}\right)^N - 1}{\left(2^{\frac{1}{N}}-1\right) N (x-1)+x y}.
\end{align}

Finally we obtain a generalised version of Model II by taking $N\to\infty$; we call this generalised because unlike the one in the main text, here for completeness we return to $u=M/N$ so that we can model replication in Model II. This yields
\begin{align}
\label{genfnNfinalIIpop}
G_{N,p}(z) &= \underset{s=0}{\overset{\infty }{\prod }}\exp\left({\frac{(1-u)2 x y}{2^s - x}}\right) \exp\left( \frac{u x y}{2^s - x}\right) = \exp\left( (2(1-u)+u)xy \sum_{s=0}^{\infty}\frac{1}{2^s-x}\right),\\
\label{genfnWfinalIIpop}
    W(z) &= \left(\log 2\right) (x-1) \left( \frac{\exp \left(\frac{(1-u) (x (2 y+\log 2)-\log 2)}{x-1}\right)\left(\exp \left(\frac{u (x ( y+\log 2)-\log 2)}{x-1}\right) -1\right)}{x( y + \log 2) - \log 2} \right. \notag \\
    &\left. \quad + \frac{\exp \left(\frac{(1-u) (x (2 y+\log 2)-\log 2)}{x-1}\right)-1}{x( 2y + \log 2) - \log 2}  \right) .
\end{align}
Note that for $u=1$ this further simplifies to
\begin{equation}
    G_{p}(z) = \frac{(x-1) \log 2 }{x y+(x-1) \log 2}\left( \exp\left(\frac{x y+(x-1) \log 2}{x-1}\right)-1\right)\exp\left( xy \sum_{s=0}^{\infty}\frac{1}{2^s-x}\right),
\end{equation}
which is the population equivalent of Model II presented in the main text. In particular we can now write the mean of protein numbers for generalised Model II, III and IV respectively
\begin{align}
\langle n \rangle_p &= \frac{2^{1-u}}{\log 2}\alpha y, \\  \langle n \rangle_p^E &= \frac{1}{N(2^{1/N}-1)}\alpha y,\\
    \langle n \rangle_p^H &= \frac{2^{1-(M/N)}}{N(2^{1/N}-1)}\alpha y \label{eq:meanIVpop},
\end{align}
where we recall $u=M/N$ in the limit of $N\to\infty$. We can also write expressions for the variances for generalised Models II, III and IV
\begin{align}
    \sigma^2_p &= \frac{2^{1-u}}{\log 2}\alpha y + \frac{4\left(u \log 2 -2\log 2 + 2^{-u} 3 \right)}{3 \log 2}\alpha^2 y  \notag \\
    &+ \frac{ 2^{1-2 u} \left(-4^u \left(u^2 \log ^2 2-u 2\log 2 (1+2 \log
   2)+2+4 \log ^2(2)+4\log 2\right)+2^{u+1} (3+2\log 2))-2\right)}{\log^2 2}\alpha^2 y^2, \\
   \left(\sigma^E_p\right)^2 &= \frac{1}{N(2^{1/N}-1)}\alpha y + \left(\frac{2}{\left(2^{\frac{1}{N}}-1\right) N}-\frac{1}{3} 2^{2-\frac{1}{N}} \right) \alpha^2 y  + \left(\frac{1}{\left(2^{\frac{1}{N}}-1\right)^2 N^2}-\frac{2^{1-\frac{2}{N}} (3 N+1)}{3
   N}\right) \alpha^2 y^2, \\
   \left(\sigma^H_p\right)^2 &= \frac{2^{1-(M/N)}}{N(2^{1/N}-1)}\alpha y + \frac{4 \left(3\cdot 2^{-\frac{M}{N}}+2^{-1/N} (2 N-M)+M-2 N\right)}{3
  \left(2^{\frac{1}{N}}-1\right) N} \alpha^2 y \notag \\
    & + \left(    - \frac{2^{1-\frac{2}{N}} 
    \left(3 \left(2^{\frac{1}{N}}-1\right) \left(-M^2+2^{\frac{1}{N}} \left(-4 M N+(M-3)
  M+4 N^2\right)+4 M N+M-4 N^2\right) \right)}
  {3  \left( 2^{\frac{1}{N}}-1 \right)^2 N^2} \right. \notag \\
  &\qquad \left. + \frac{  \left(4^{\frac{M}{N}}\left( -5\ 2^{\frac{1}{N}+2} N+4
  N+2^{\frac{N+2}{N}} (8 N+3)\right)+3 \left(2^{\frac{1}{N}} (2 N+3)-2 N\right)
  2^{\frac{M+N+1}{N}}-3\ 2^{\frac{N+2}{N}}\right)}{3
  \left(2^{\frac{1}{N}}-1\right)^2 N^2}  \right) \alpha^2 y^2\label{eq:varIVpop}.
\end{align}

Note that generalised Model II for $u=1$ simplifies to Model II in the main text
\begin{align*}
\langle n \rangle_p = \frac{\alpha y}{\log 2}, \ \sigma^2_p = \frac{1}{\log 2}\alpha y + \frac{6 - 4\log 2}{3\log 2}\alpha^2 y + \frac{1-\log^2 2}{\log^2 2} \alpha^2 y^2 \simeq 1.44\alpha y +1.55\alpha^2 y + 0.08\alpha^2 y^2.
\end{align*}

\bibliographystyle{unsrt}
\bibliography{CasperPaper}

\newpage

\begin{figure}
\centering
\subfloat[]{
    \includegraphics[height=68mm]{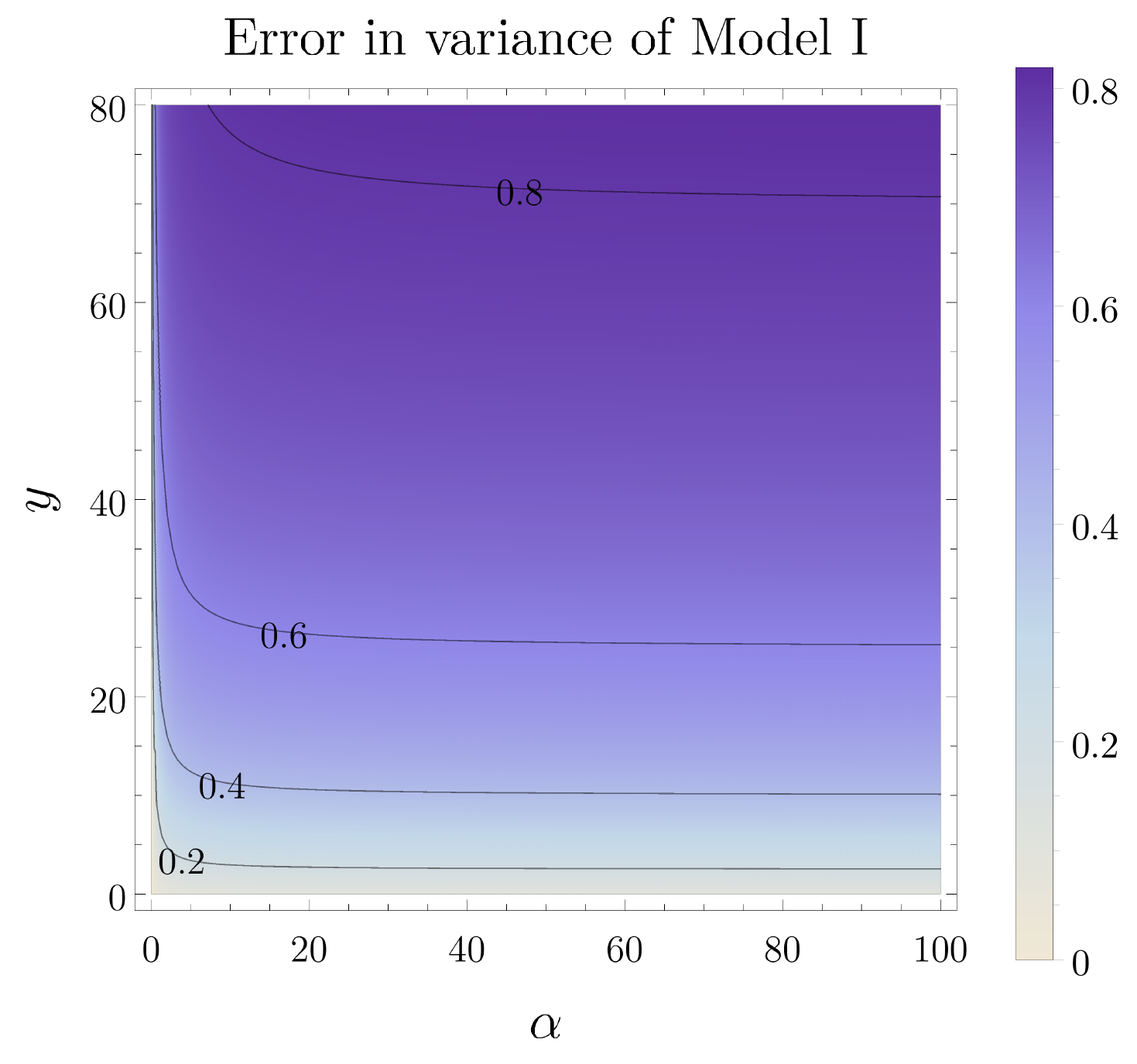}
}
\subfloat[]{
    \includegraphics[height=68mm]{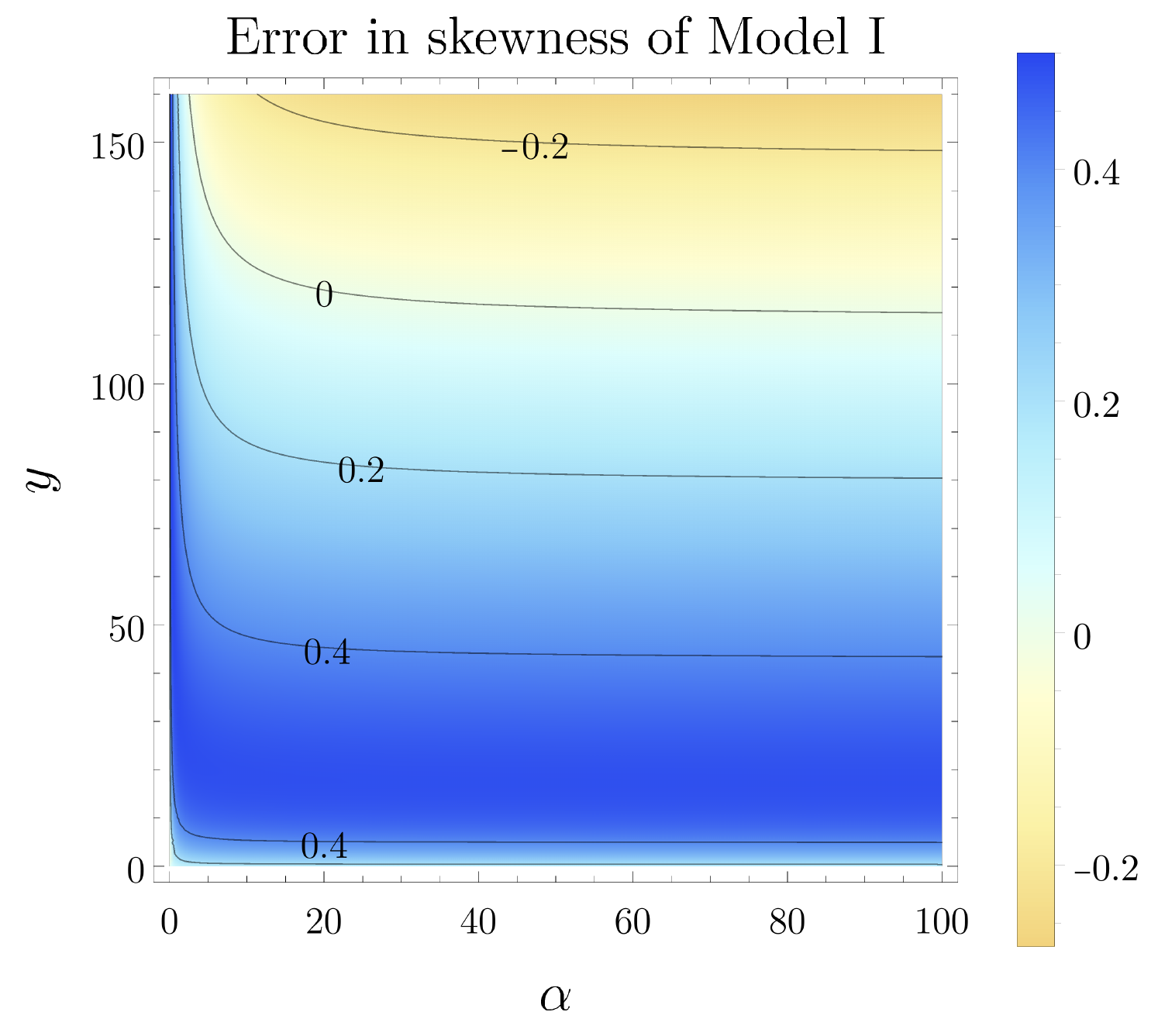}
}
\caption{(a) Plot of the relative error between the steady-state variance of Model I and the cyclo-stationary variance of Model II as a function of the mean burst size ($\alpha$) and the mean number of mRNAs produced in one cell cycle ($y$), assuming lineage measurements. (b) Plot of the relative error between the skewness of Model I and II as a function of $\alpha$ and $y$. Note that for burst sizes $\alpha \gtrapprox 10$, the relative errors are a strong function of $y$, with only a weak dependence on $\alpha$. Model I in all cases underestimates the variance of Model II. In contrast, Model I underestimates the skewness of Model II for small $y$ and overestimates it for large $y$. The relative error in the variance is defined as $(\sigma^2_{s}-\sigma^2_{NB,s})/\sigma^2_{s}$ and the relative error in the skewness is defined as $(S_{s}-S_{NB,s})/S_{s}$.}
\end{figure}

\begin{figure}[h]
\centering
\subfloat[$\alpha = 1, y=5$]{
  \includegraphics[width=75mm]{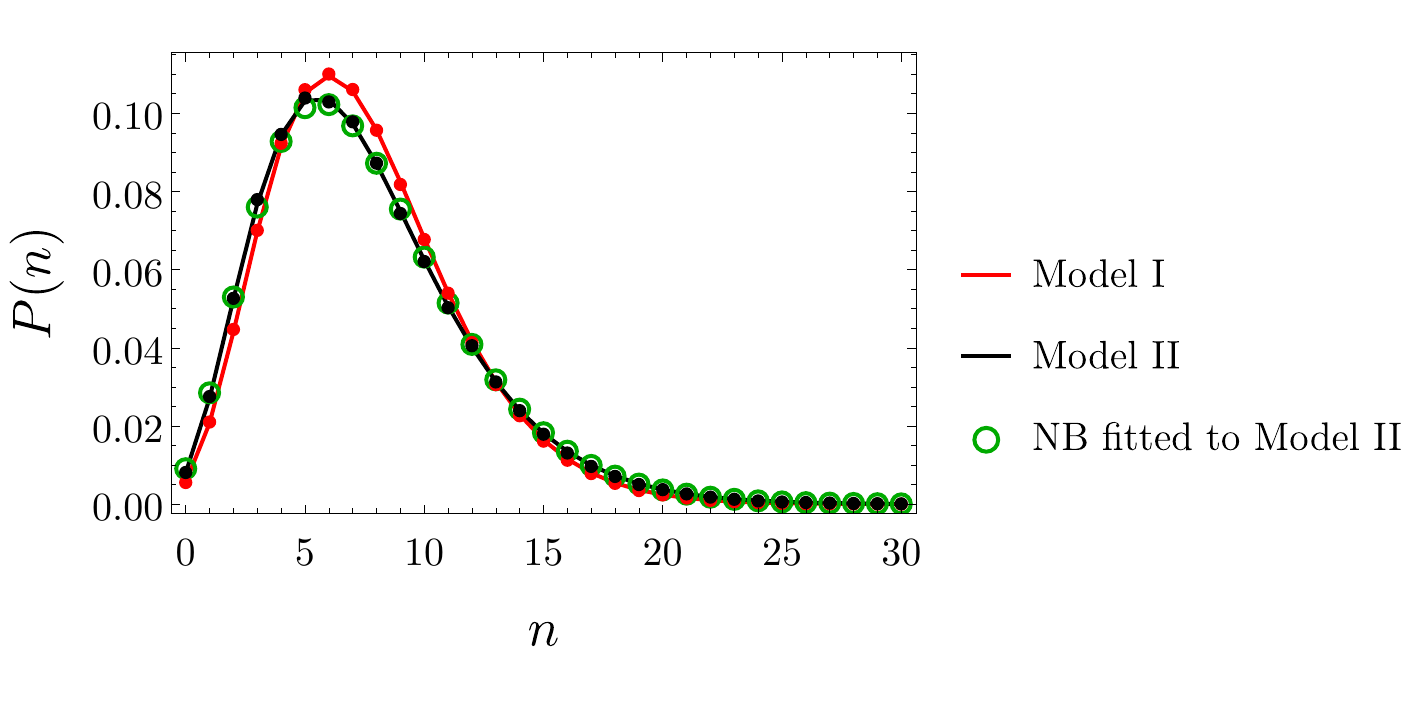}
}
\subfloat[$\alpha = 1, y=20$]{
    \includegraphics[width=75mm]{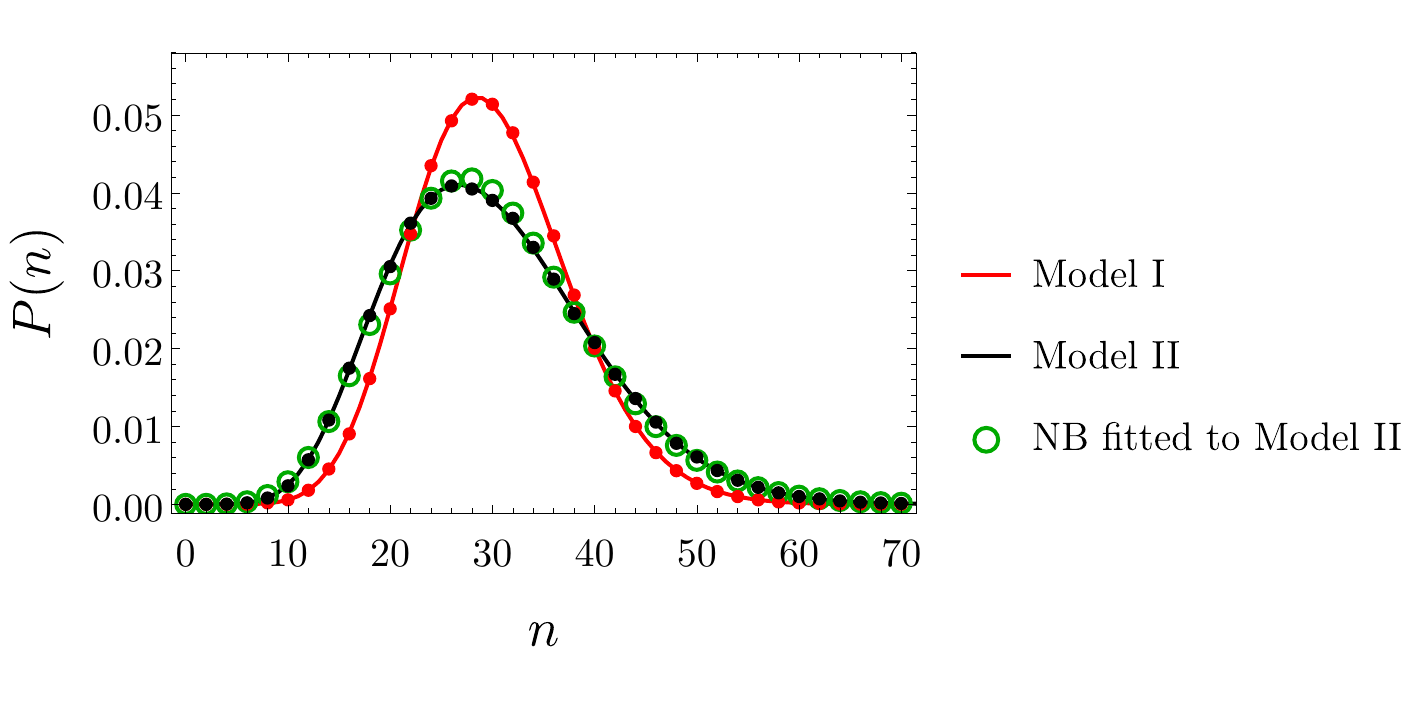}
}
\hspace{0mm}
\subfloat[$\alpha = 1, y=40$]{
  \includegraphics[width=75mm]{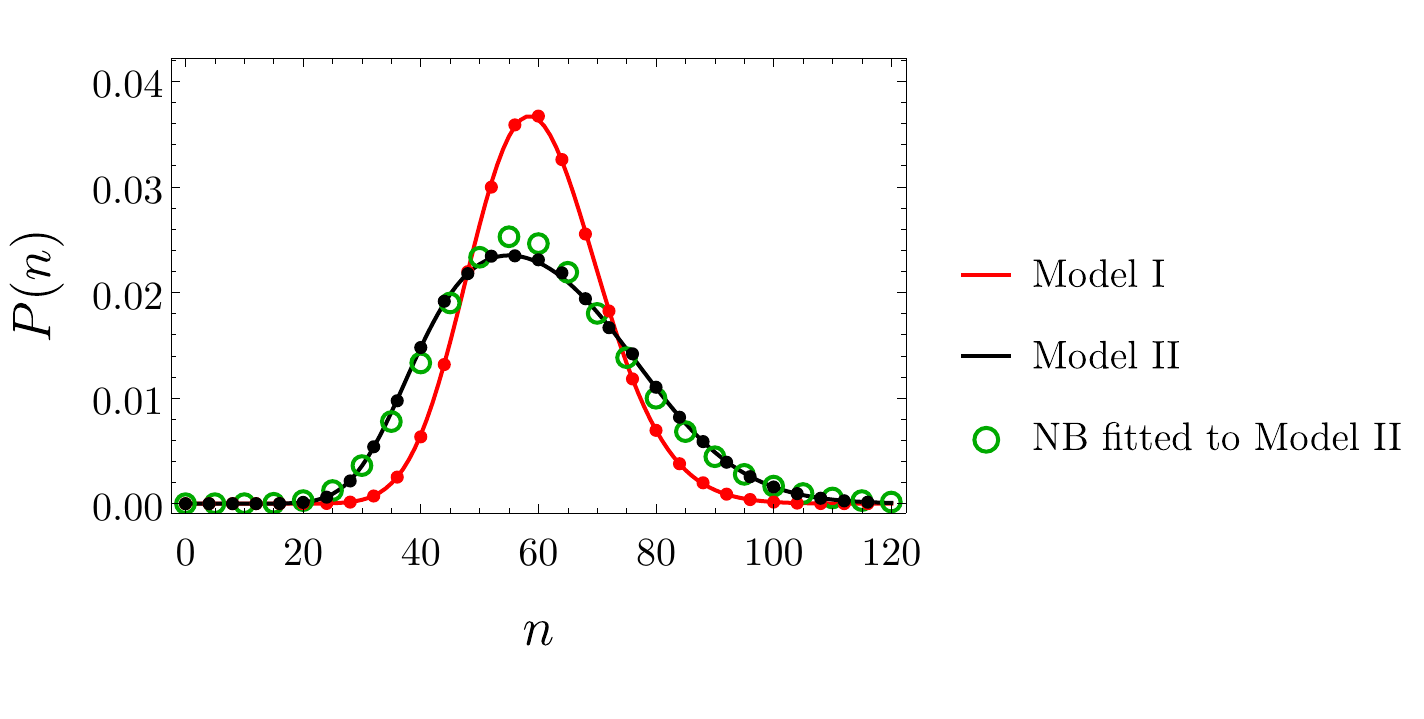}
}
\subfloat[$\alpha = 1, y=100$]{
  \includegraphics[width=75mm]{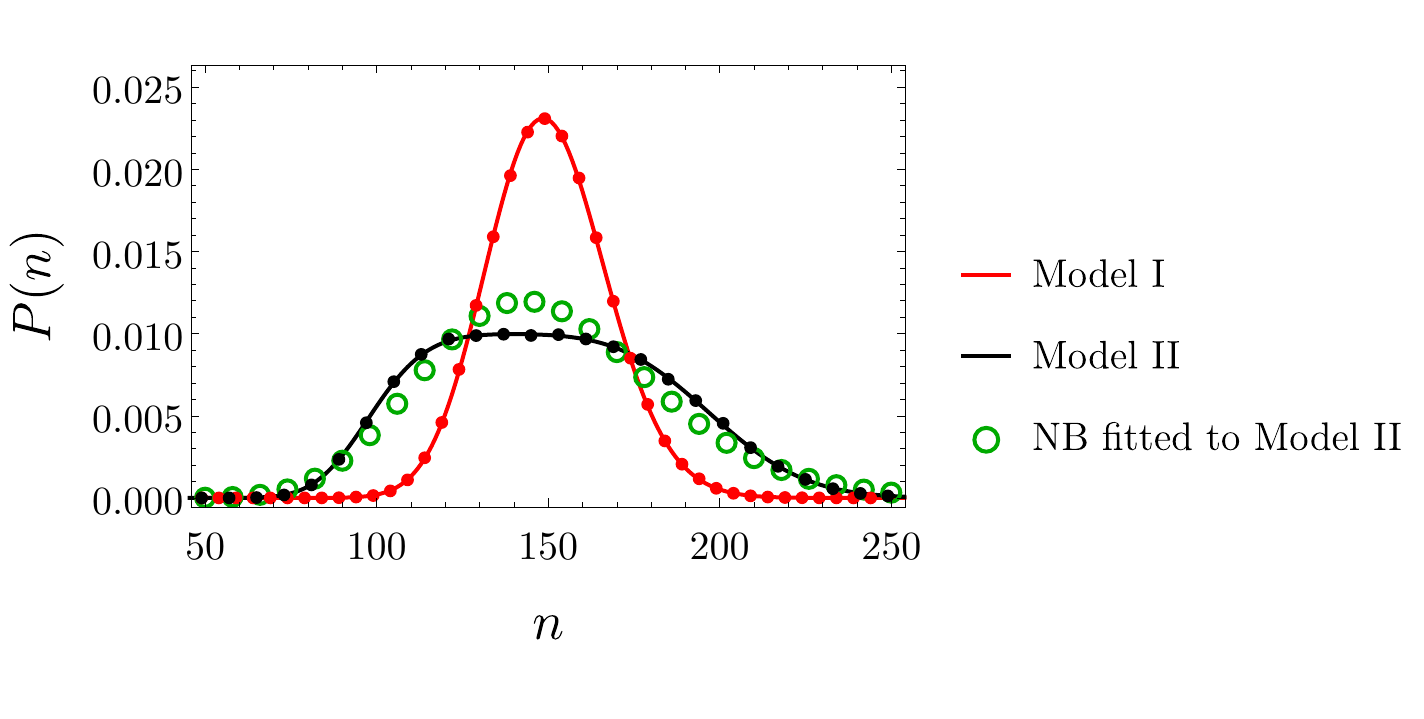}
  
}
\caption{Comparison of the protein number distributions (assuming lineage measurements) of Model I and Model II in steady-state and cyclo-stationary conditions, respectively. For Model I (red line), the distribution is $\text{NB}(3y/2,\alpha/(1+\alpha))$ and for Model II (black line) it is given by $P(n)=(1/n!)dG_s^n/dz^n|_{z=0}$ where $G_s$ is given by Eq.\ \eqref{genfnsingle}. Note that the numerical computation of the distribution from Eq.\ \eqref{genfnsingle} can be greatly accelerated (whilst maintaining accuracy) if the sum over $s$ is truncated to a few tens of terms. The dots show the distributions of Model I and Model II obtained from stochastic simulations (simulations of Model I are done using the conventional SSA and those of Model II are done using a modified SSA, see Appendix B and also at the end of this caption). The open green circles show the negative binomial distribution which has the same first and second moments as the distribution of Model II. The stochastic simulations of Model II are performed as follows. Initially we have a single cell with zero protein. We measure the protein content of the cell at intervals $T/Z$ where $Z = 10 \pi$. Each time a cell divides, we follow only one of the daughter cells. The simulation is run until $10^5$ cycles have passed and a histogram is calculated from this data (we discard the first $10^3$ cell cycles to ignore any possible transients). The cell cycle length is $T = 1$ in all cases. All lineage simulations in this article use this protocol, unless otherwise stated.
}
\end{figure}

\begin{figure}[h]
\centering
\includegraphics[width=130mm]{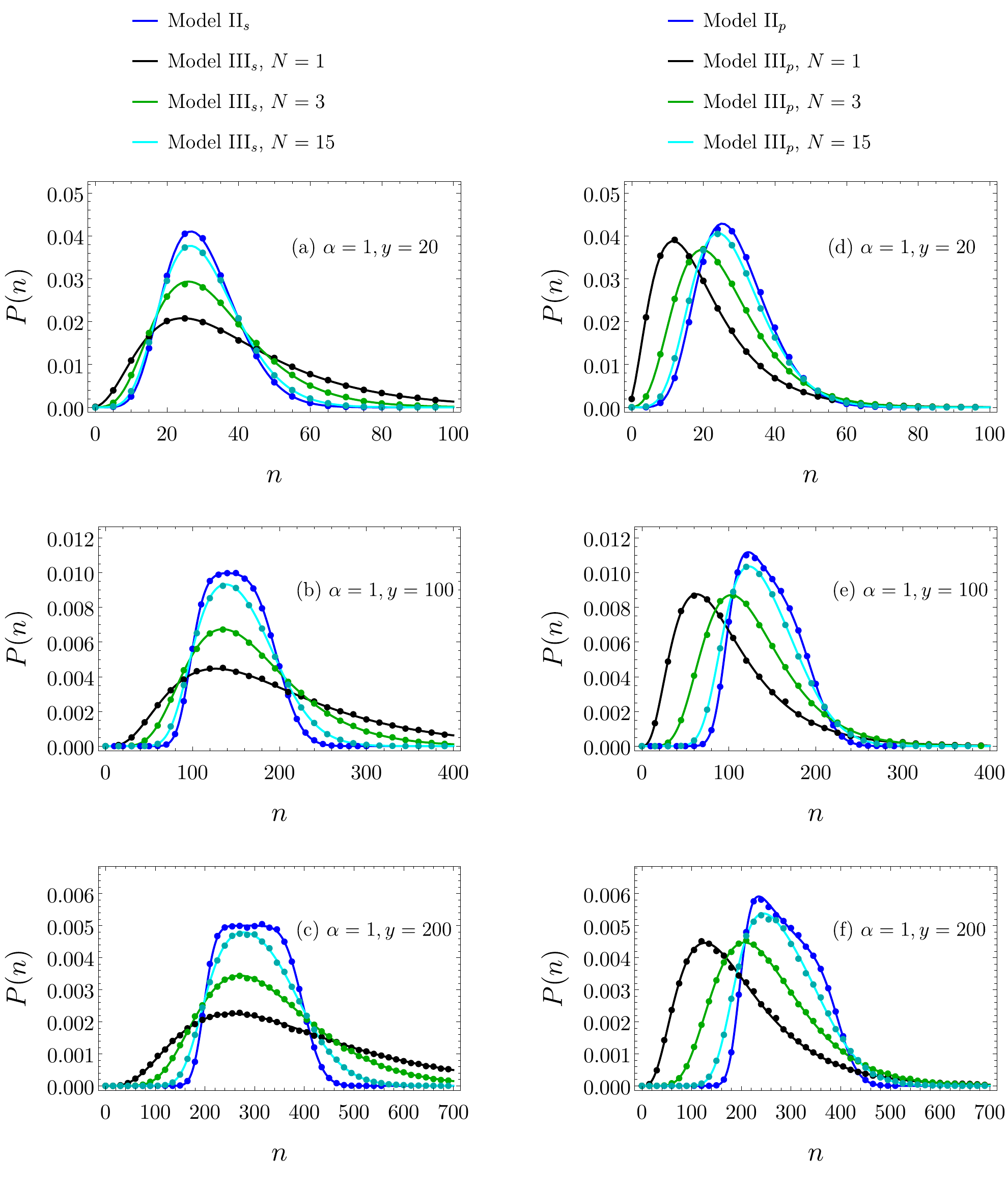}
\caption{Plots of the protein number distribution $P(n)$ 
for Model II (blue line) and Model III (black, green, cyan lines) with single lineage (a-c) and population observations (d-f). The solid dots show the distributions obtained from stochastic simulations using the SSA (see Appendix B) which agree with those from theory in all cases. Note that in the limit of large number of cycle cycle phases $N \gg 1$, Model III approaches Model II since the cell cycle length variability tends to zero. The non negative binomial nature of Model II for large $y$ (the flat region near the mode in b and c and the right shoulder in e and f) is washed away as the cell cycle length variability increases, i.e.\ as $N$ decreases in Model III. Note that the numerical computation of the distribution of Model III from Eqs.\ \eqref{genfnErlangfinal1} and \eqref{genfnErlangfinal1pop} can be greatly accelerated (whilst maintaining accuracy) if the infinite product is truncated to a few tens of terms.
}
\end{figure}

\begin{figure} [h]
\centering
    \includegraphics[width=120mm]{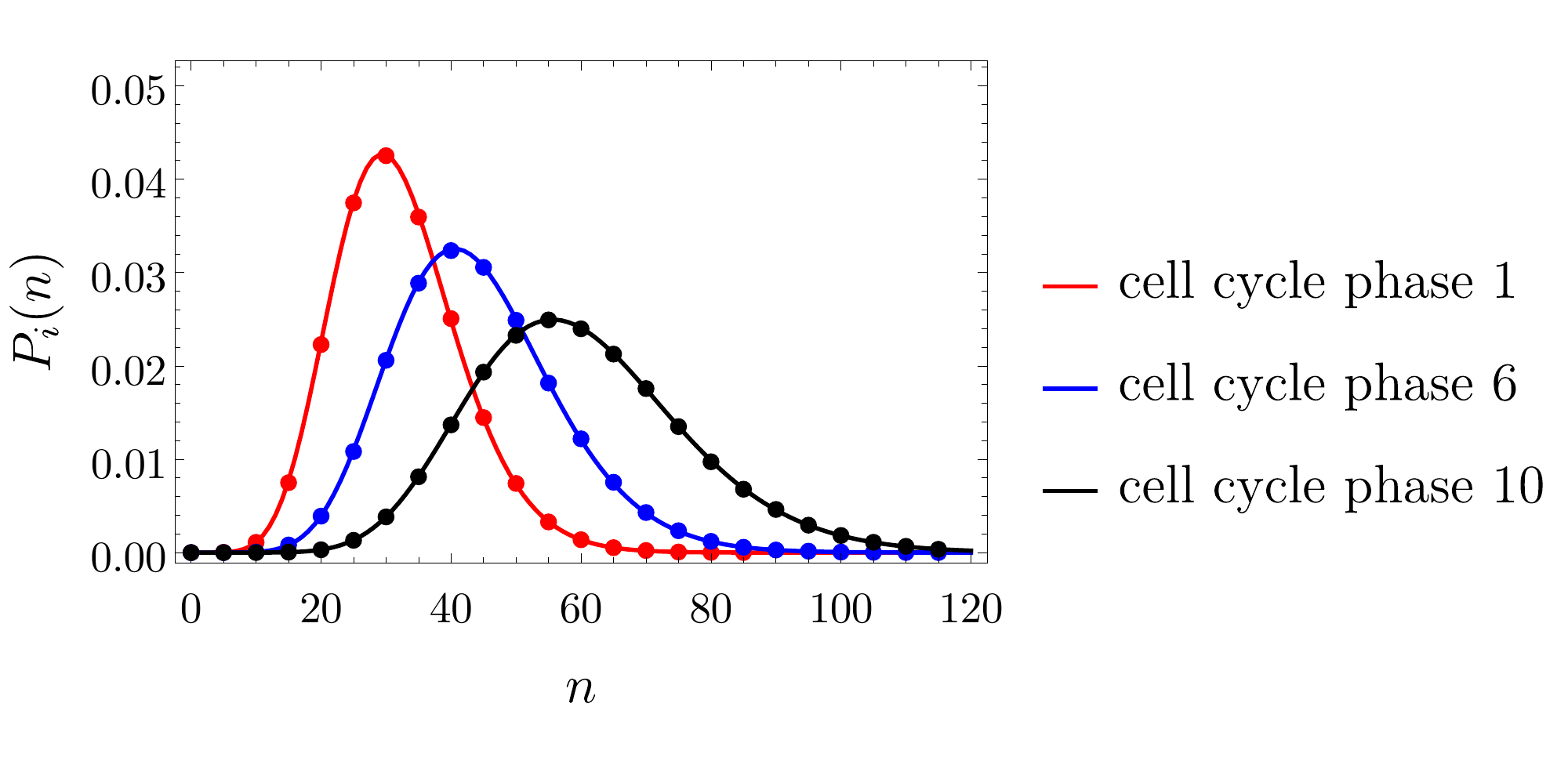}
\caption{Plots of the protein distribution at different points in the cell cycle for Model IV with an Erlang distributed cell cycle (10 phases) and replication occurring at the start of phase 6. The distributions (solid lines) are calculated from the theory for lineage observations while the dots show the same calculated from a single trajectory generated by the SSA for $10^6$ cell cycles. Note that the theory is for Model IV Eq.\ \eqref{finalgenfnhypoexp_i} with $i=1,6,10$, $k_i = k = N/T$ and $\alpha_i=\alpha$ for all $i$, and $r_i = r, \ i \in [1,M]$ and $r_i = 2r, \ i \in [M+1,N]$, where $M$ is the cell cycle stage in which replication occurs. The parameters are $N=10, M=5,\alpha=1,y=20,T=1$. 
}
\end{figure}

\begin{figure} [h]
\centering
\subfloat[]{
\includegraphics[width=125mm]{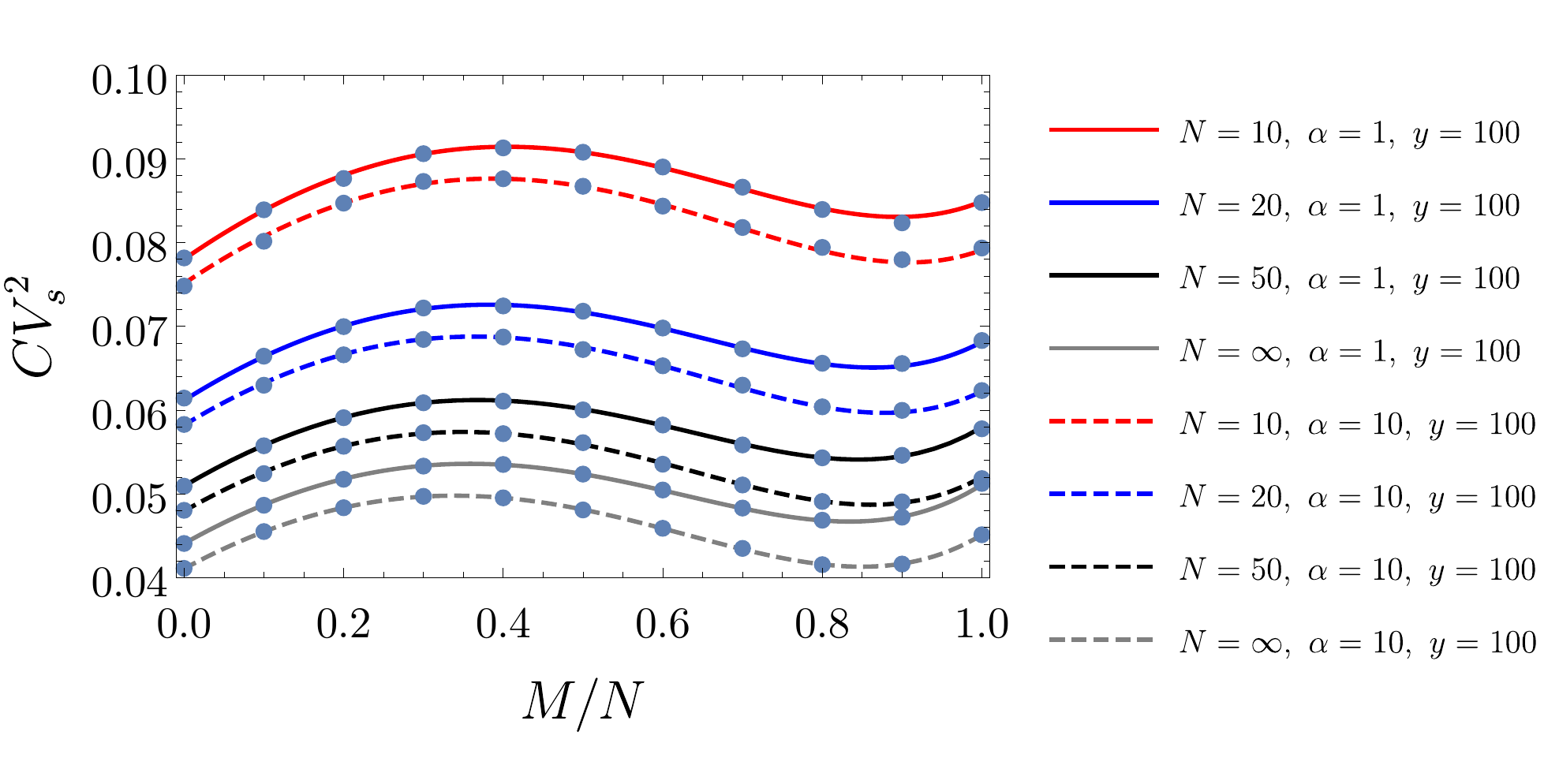}
}
\\
\subfloat[]{
\includegraphics[width=125mm]{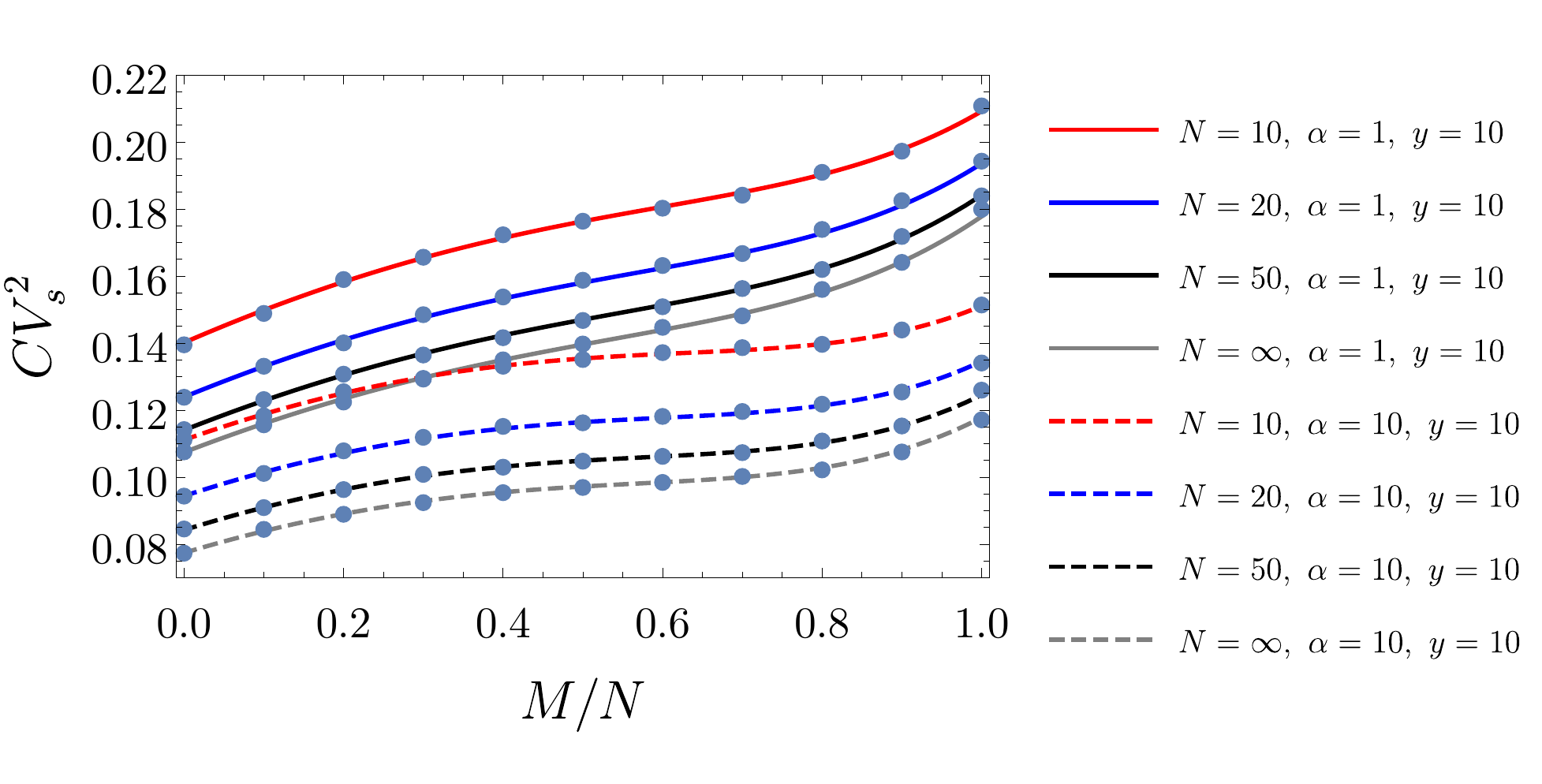}
}
\caption{Effects of the position of the replication point in the cell cycle on the size of protein fluctuations for lineage observations. Plots of the coefficient of variation squared $CV_s^2$ as a function of the mean burst size $\alpha$, the mean number of mRNAs produced in a cell cycle $y$ and the fraction of the cell cycle in the pre-replication stage $M/N$ for Model IV (same setup as previous figure). The mean and variance for the computation of the $CV_s^2$ are given by Eqs.\ \eqref{meanmodelIVsp} and \eqref{varmodelIVsp}. A comparison of (a) and (b) shows that the $CV_s^2$ decreases with increasing $\alpha$ and $y$. However the $CV_s^2$ has a complex dependence on $M$: for large $y$, $CV_s^2$ is roughly independent of $M$ while it increases with $M$ for small $y$. Solid dots show the results of the SSA and solid lines show the theory. 
}
\end{figure}

\begin{figure} [h]
\centering
\includegraphics[width=130mm]{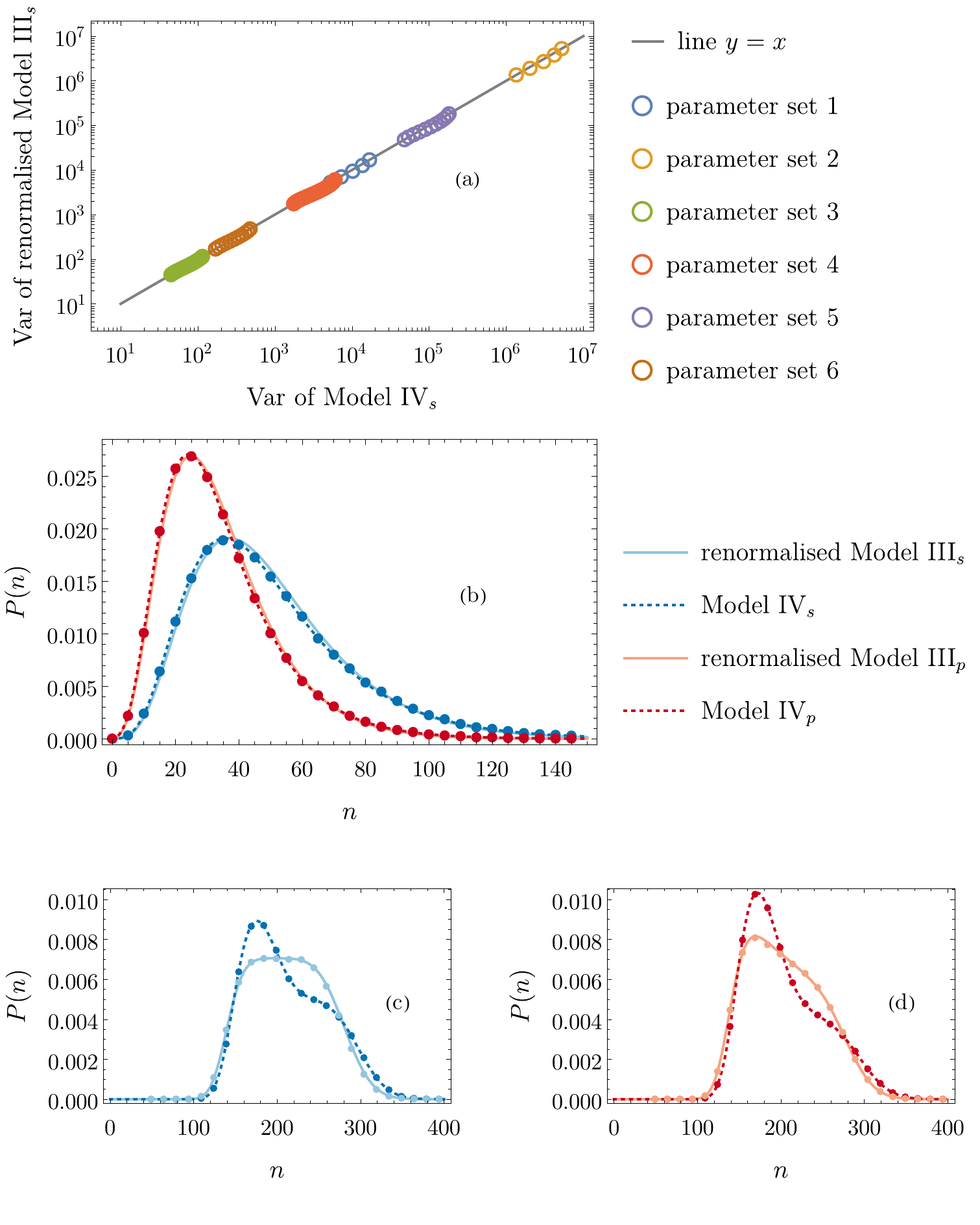}
\caption{Relationship between Model III and Model IV for lineage (denoted by subscript $s$ in figure) and population observations (denoted by subscript $p$ in figure). (a) renormalising the parameter $y$ in Model III (lineage) by changing it to $\Lambda_s y$, we find that the variance agrees to a good degree of approximation with the variance of Model IV (lineage) for all parameter sets tested. The mean of the two models is exactly the same under this renormalisation. The same applies for population observations (not shown). (b) The protein distribution of Model IV and of the renormalised Model III are also in excellent agreement (these are obtained from the PGF solutions of both models) for small $y$ and $N$. Note the renormalisation factors are given by Eq.\ \eqref{lambdadefinition} for lineage and $\Lambda_p=2^{1-M/N}$ for population observations. (c) In contrast the distributions of Model IV (lineage) and of renormalised Model III (lineage) are very different when $y$ and $N$ are large. In (d) we show that the same conclusion holds for population observations. In all cases, dots show the distribution obtained from the SSA is in good agreement with the theory (solid or dashed lines) for Models III and IV. Parameter sets are as follows. The 6 parameter sets for Model IV in (a) are: (1) $N=4, \alpha=10, y = 10$, (2) $N=4, \alpha=20, y = 100$, (3) $N=20, \alpha=1, y = 10$, (4) $N=20, \alpha=2, y = 50$, (5) $N=10, \alpha=5, y = 100$, and (6) $N=10, \alpha=2, y = 10$; for renormalised Model III, the parameters are the same except that $y$ is renormalised. For each parameter set we take $M=0,\dots,N$. Parameters for (b) are $N=2,\alpha = 1.0,y = 20, M = 1$ for Model IV and same but with $y = 20 \Lambda_{s,p}$ for renormalised Model III. Parameters for (c, d) are $N\to\infty,\alpha=1.0,y=100,M=0.5N$ for Model IV and same for renormalised Model III, but with $y=100\Lambda_{s,p}$. Population snapshot SSA data for Model IV consists of $\sim 10^6$ cells starting from a single cell with zero protein content (see Appendix B for details of the simulations).
}
\end{figure}

\begin{figure} [h]
\centering
\subfloat[]{
  \includegraphics[width=110mm]{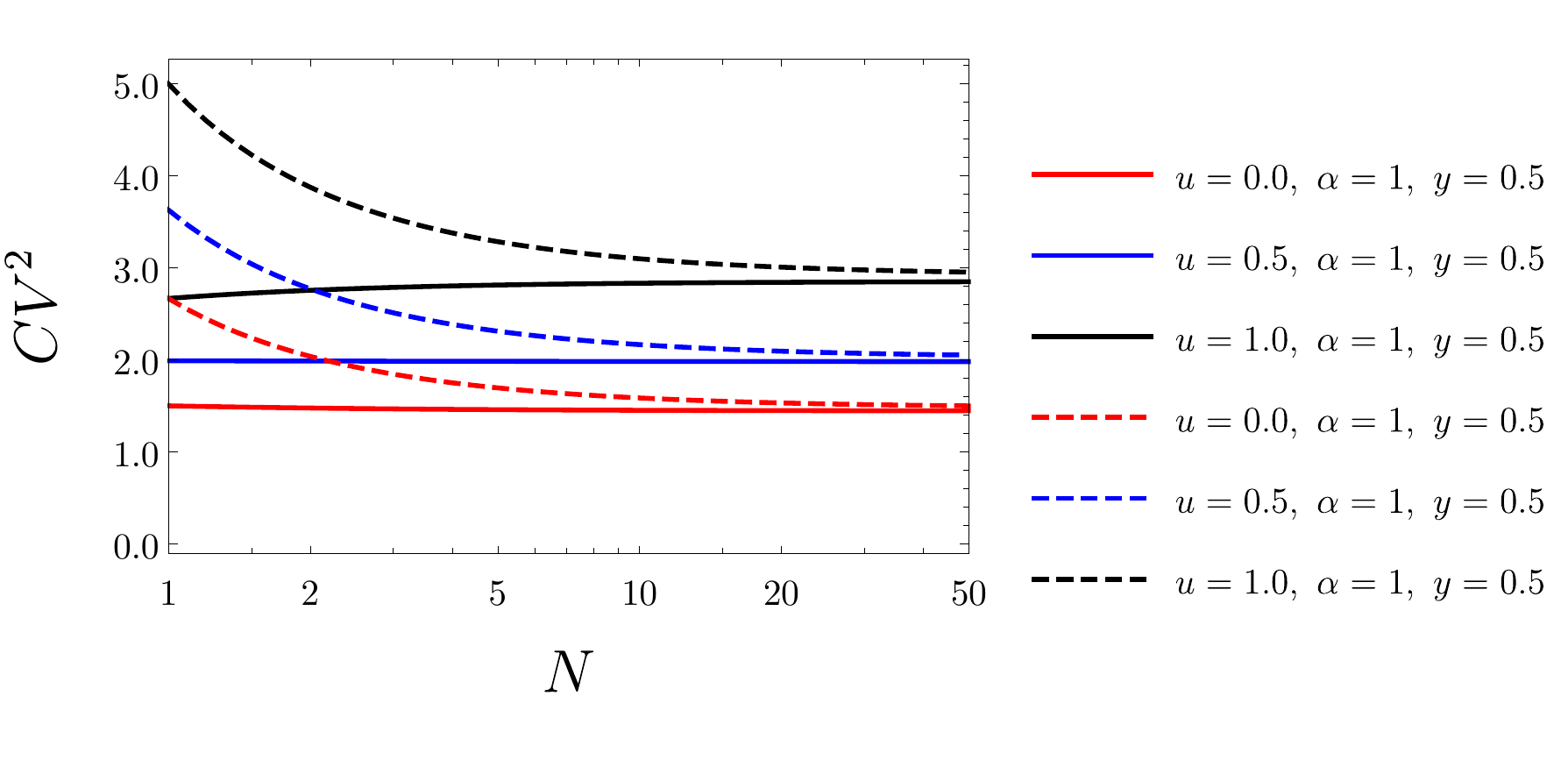}
}
\\
\subfloat[]{
\includegraphics[width=110mm]{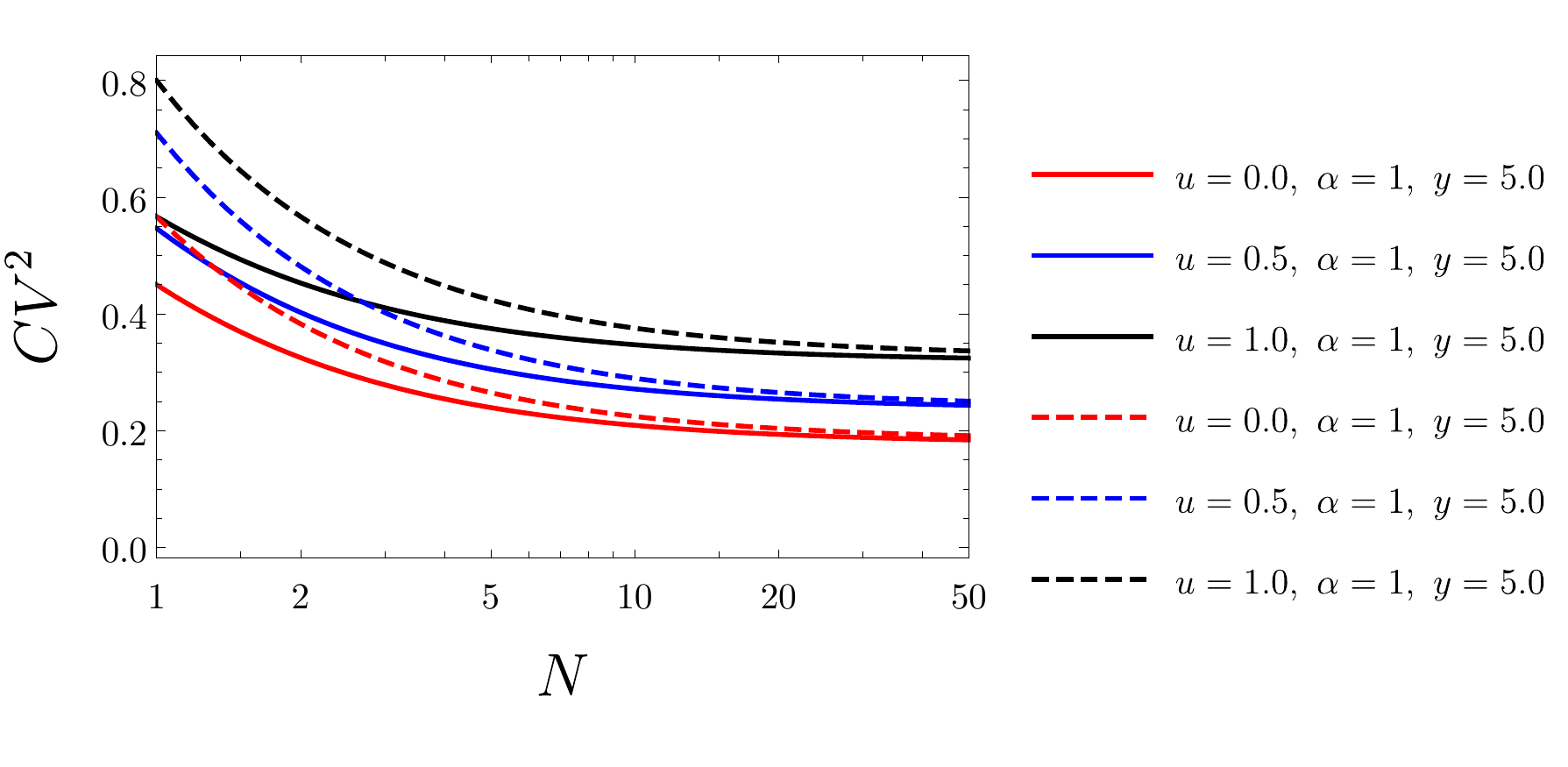}
}
\\
\subfloat[]{
\includegraphics[width=110mm]{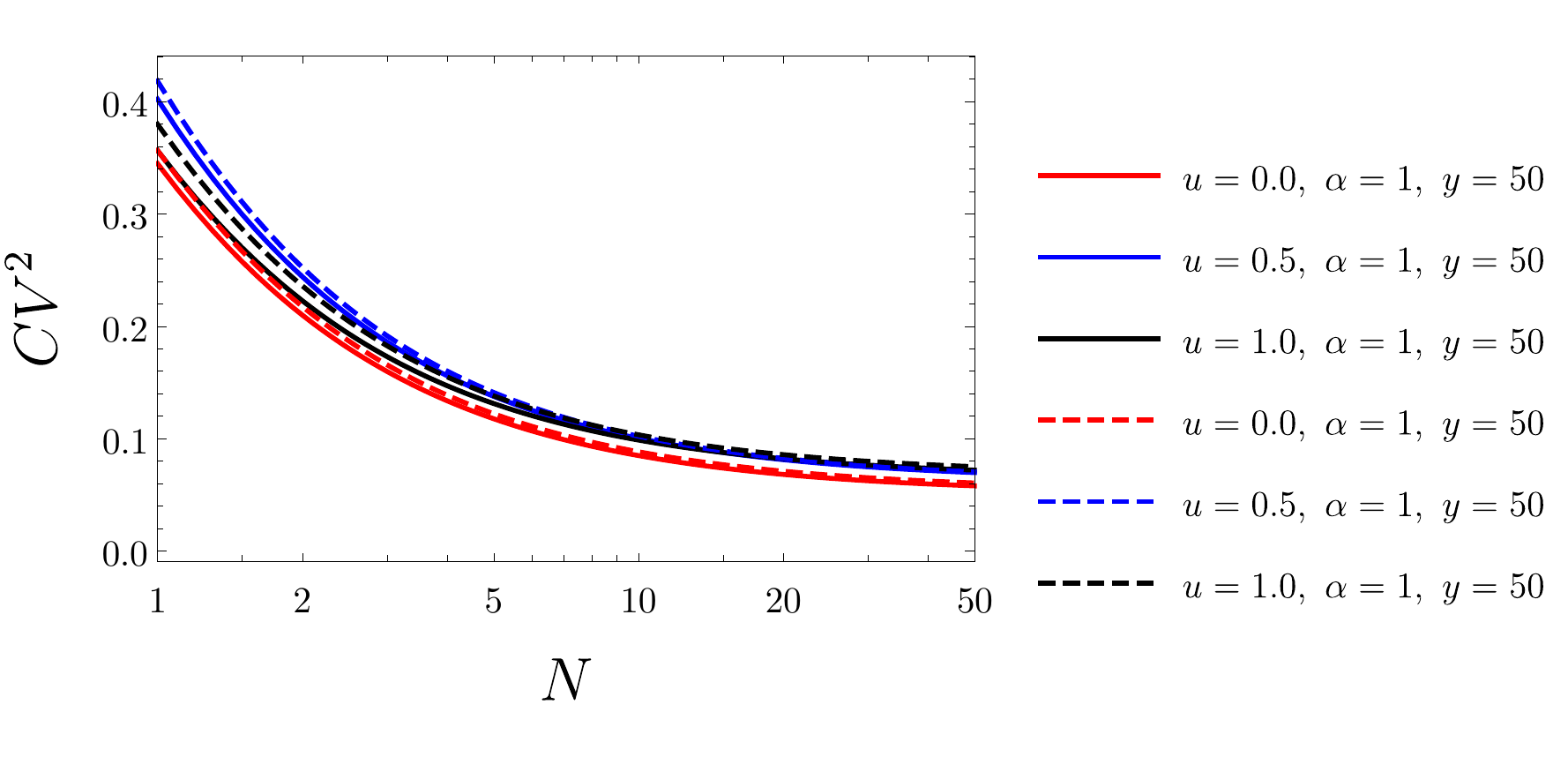}
}
\caption{Effects of the cell cycle variability on the size of protein fluctuations showing the difference between lineage (solid) and population snapshot (dotted) observations. Plots of the coefficient of variation squared $CV^2$ as a function of the mean burst size $\alpha$, the mean number of mRNAs produced in a cell cycle $y$ and the fraction of the cell cycle in the pre-replication stage $u=M/N$. The mean and variance for the computation of the $CV_s^2$ are given by Eqs.\ \eqref{meanmodelIVsp} and \eqref{varmodelIVsp}, whereas for $CV_p^2$ we have used Eqs.\ \eqref{eq:meanIVpop} and \eqref{eq:varIVpop}. A comparison of (a), (b) and (c) shows that the $CV_s^2$ can be both increasing and decreasing with increasing $N$, but $CV_p^2$ is always decreasing with increasing $N$, confirming theoretical predictions. For large values of $y$, we see that the protein noise only weakly depends on $u$, which can be understood from noting that this is the regime in which protein noise is dominated by the extrinsic noise floor. 
}
\end{figure}

\end{document}